\documentstyle[11pt]{article} 
\def\esp#1{} \textwidth=6.5truein \textheight=9truein \hoffset=-1.5cm
\voffset=-2cm \font\shell=msym10 \def\C{\mbox{\shell C}}
\def\Cdiff{\C_{\mbox{\scriptsize diff}}} \def\q{\quad} \def\qq{\qquad}
\def\d{\mbox{d}} \def\i{\mbox{i}} \def\e{\mbox{e}} \def\p{Pain\-lev\'e}
\def\tx{\textstyle} \def\tfr#1#2{{\tx{#1\over#2}}}

\def\v#1{\mbox{\bf v}_{#1}}
\def\~#1{\mbox{\protect$\bf\tilde{\mit#1}$}} 
\def\N{\mbox{\shell N}} \def\e{\mbox{e}} 
\def\al{\alpha} \def\be{\beta} \def\pa{\partial}
\def\and{\quad\mbox{and}\quad}

\def\sqr#1#2{{\vcenter{\vbox{\hrule height.#2pt
	      \hbox{\vrule width.#2pt height#1pt \kern#1pt \vrule
	      width.#2pt} \hrule height.#2pt}}}}
\def\square{\mathchoice\sqr34\sqr34\sqr{2.1}3\sqr{1.5}3}
 \def\i{\mbox{i}}
\def\fu{f_u} \def\fuu{f_{uu}}
\def\fuuu{f_{uuu}}
\def\fuuuu{f_{uuuu}}
\def\fuuuuu{f_{uuuuu}}
\def\ex#1{\exp\left\{#1\right\}}

  \def\DGBS{{\sc dgbs}}
\def\DGB{{\sc dgb}} \def\PDE{{\sc pde}} \def\PDES{{\sc pdes}}
\def\ODE{{\sc ode}} \def\ODES{{\sc odes}}
\def\bigskip{\vspace{\bigskipamount}}
\def\medskip{\vspace{\medskipamount}}
\def\smallskip{\vspace{\smallskipamount}}

\def\cc#1{\kappa_{#1}} \newcommand{\beq}{\begin{equation}}
\newcommand{\beqn}{\begin{displaymath}}
\newcommand{\bear}{\begin{eqnarray} \baselineskip=12pt}
\newcommand{\bearn}{\begin{eqnarray*} \baselineskip=12pt}
\newcommand{\eeq}{\end{equation}} \newcommand{\eeqn}{\end{displaymath}
\noindent} \newcommand{\eear}{\end{eqnarray}}
\newcommand{\eearn}{\end{eqnarray*}}

\newcommand{\newsection}{\setcounter{equation}{0}
	    \section}

\begin{document} \begin{titlepage} \pagestyle{empty}
\esp{\begin{frontmatter}} \title{Symmetry Reductions and Exact
Solutions \\
 of a class of Nonlinear Heat Equations} \author{ \vspace{.5cm} by \\
\vspace{.5cm} {\sc Peter A. Clarkson and Elizabeth L. Mansfield} \\
\vspace{.5cm} \\ Department of Mathematics, University of Exeter,
Exeter, EX4 4QE, U.K.\\ and\\ Program in Applied Mathematics,
University of Colorado,\\ Boulder, CO 80309-0526, U.S.A.} \maketitle
\vspace{1in} \begin{abstract}

Classical and nonclassical symmetries of the nonlinear heat equation
$$u_t=u_{xx}+f(u),\eqno(1)$$ are considered. The method of differential
Gr\"obner bases is used both to find the conditions on $f(u)$ under
which symmetries other than the trivial spatial and temporal
translational symmetries exist, and to solve the determining equations
for the infinitesimals. A catalogue of symmetry reductions is given
including some new reductions for the linear heat equation and a
catalogue of exact solutions of (1) for cubic $f(u)$ in terms of the
roots of $f(u)=0$.  \end{abstract} \esp{\end{frontmatter}}
\end{titlepage}

\newsection{Introduction}

The nonlinear heat equation \beq u_t = u_{xx}+f(u), \label{nheateq}\eeq
where $x$ and $t$ are the independent variables, $f(u)$ is an arbitrary
sufficiently differentiable function and subscripts denote partial
derivatives, arises in several important physical applications
including microwave heating (where $f(u)$ is the rate of absorption of
microwave energy, cf.\ \cite{PinSm,Smyth}), in the theory of chemical
reactions (where $f(u)$ is the temperature dependent reaction rate,
cf.\ \cite{Ames:67,Aris,Frank}) and in mathematical biology (where
$f(u)$ represents the reaction kinetics in a diffusion process, cf.\
\cite{Murray}).

The classical method for finding symmetry reductions of \PDES\ is the
Lie group method of infin\-ites\-imal transformations
\cite{Ames,BCb,BK,Hill,Olver,Ovs,RAmes,Step}.  To apply the classical
method to the nonlinear heat equation (\ref{nheateq}) we consider the
one-parameter Lie group of infinitesimal transformations in $(x,t,u)$
given by \bear \~{x} &=&{x}+ \varepsilon {\xi}(x,t,u) +
O(\varepsilon^2),
 \nonumber \\ \~{t} &=& {t} + \varepsilon {\tau}(x,t,u) +
O(\varepsilon^2),
 \\ \~{u} &=& {u} + \varepsilon {\phi}(x,t,u) + O(\varepsilon^2),
 \nonumber \eear where $\varepsilon$ is the group parameter.  Requiring
that (1.1) is invariant under this transformation yields an
overdetermined, linear system of equations for the infin\-ites\-imals
${\xi}(x,t,u)$, ${\tau}(x,t,u)$ and ${\phi}(x,t,u)$.  The associated
Lie algebra of infin\-ites\-imal symmetries is the set of vector fields
of the form \beq {\v{}} = \xi(x,t,u){\partial\over\partial x} +
\tau(x,t,u){\partial\over\partial t}
 +\phi(x,t,u){\partial\over\partial u}.\eeq Classical symmetries of the
nonlinear heat equation (1.1) have studied by Dorodnitsyn \cite{Dorod}
who classified conditions on $f(u)$ for which symmetries exist, though
he did not give the associated reductions (see also \cite{Frey,OrR}).

There have been several generalizations of the classical Lie group
method for symmetry reductions. Ovsiannikov \cite{Ovs} developed the
meth\-od of partially invariant solutions.  Bluman and Cole \cite{BCa},
in their study of symmetry reductions of the linear heat equation,
proposed the so-called nonclassical method of group-invariant solutions
(in the sequel referred to as the {\it nonclassical method\/}), which
is also known as the ``method of conditional symmetries''
\cite{Fush,FNik,FZh,Gaeta,LWint,ORb,Pucci,PS,Wint} and the ``method of
partial symmetries of the first type'' \cite{Vor}. In this method, the
original \PDE\ (1.1) is augmented with the invariant surface condition
\beq \psi\equiv\xi(x,t,u)u_x + \tau(x,t,u)u_t -\phi(x,t,u)=0,\eeq which
is associated with the vector field (1.3). By requiring that both (1.1)
and (1.4) are invariant under the transformation (1.2) one obtains an
overdetermined, nonlinear system of equations for the
{infin\-ites\-imals} ${\xi}(x,t,u)$, ${\tau}(x,t,u)$ and
${\phi}(x,t,u)$, which appear in both the transformation (1.2) and the
supplementary condition (1.4). Since the number of determining
equations arising in the nonclassical method is smaller than for the
classical method and since all solutions of the classical determining
equations necessarily satisfy the nonclassical determining equations,
the solution set may be larger in the nonclassical case.  For some
equations, such as the Korteweg-de Vries equation, the infinitesimals
arising from the classical and nonclassical methods coincide. It should
be emphasized that the associated nonclassical vector fields do not
form a vector space, still less a Lie algebra, since the invariant
surface condition (1.4) depends upon the particular reduction.
Subsequently, these methods were further generalized by Olver and
Rosenau \cite{ORa,ORb} to include ``weak symmetries'' and, even more
generally, ``side conditions'' or ``differential constraints'', and
they concluded that ``the unifying theme behind finding special
solutions of \PDES\ is not, as is commonly supposed, group theory, but
rather the more analytic subject of overdetermined systems of \PDES''.

In this paper we consider both classical and nonclassical symmetries of
the nonlinear heat equation (1.1), and find conditions on $f(u)$ under
which symmetries, other than trivial translational ones, may exist.
The method used to find the necessary conditions on $f(u)$, and from
there to find the symmetries, is that of differential Gr\"obner bases
(\DGBS) \cite{Mans,MF}.  This method yields the ``triangulation" of a
nonlinear system of \PDES\ from which solutions can be obtained in a
systematic way.

 The basic algorithm used to generate a \DGB\ for a system of \PDES\ is
the Kolchin-Ritt algorithm, which forms an essential part of the
StandardForm algorithm \cite{Reida,Reidb}\ and the diffGbasis algorithm
\cite{MF}. It appears to have been formally written down first by
Carr\`a-Ferro \cite{CF}.  The Kolchin-Ritt algorithm has been used for
some time, intuitively if not formally, to yield integrability
conditions for linear and orthonomic systems \cite{Kol,Reida}.  With
hindsight, one can see that the integrability conditions generated by
the Janet-Riquier algorithm, which completes a system of orthonomic
\PDES\ to involutive form (cf.  for example \cite{Ja,Jb,Schc,Stor,Top}
and references therein), are produced by exactly the calculations
performed in the Kolchin-Ritt algorithm.  (It should be noted that
several authors incorrectly assume that when the Kolchin-Ritt algorithm
terminates, the output system is involutive, or passive.  A
counter-example can be found in \cite{Mans}.)  The production of
integrability conditions requires a total degree ordering on the
derivative terms. On the other hand, the elimination ideals, or
triangulation, of a system of \PDES\ requires a purely lexicographical
ordering.

Recently, the limits of the Kolchin-Ritt algorithm in calculating a
\DGB\ of a nonlinear system were found, and sufficient conditions for
obtaining a \DGB\ for a nonlinear system of \PDES\ were formulated
\cite{Mans}. The Kolchin-Ritt algorithm generates a \DGB\ only ``up to"
(in a sense to be defined) a certain set of differential coefficients,
and these coefficients must not lie in the ideal.  Proving that these
coefficients do not lie in the ideal has no algorithmic solution as
yet, and there are reasons as to why it may not have one at all.  The
condition is also one of the sufficient conditions in \cite{MF} for a
\DGB\ of a nonlinear system.  Thus far, attempts to formulate
algorithms that circumnavigate the difficulty, for example by
calculating the algebraic Gr\"obner basis of successive prolongations
of the system \cite{CF,Oliv,Pank}, lead to processes that do not
terminate in finite time. Fortunately in many cases one can prove the
condition directly, and there is a range of techniques that can be used
when the condition fails.  Despite the limitation, the Kolchin-Ritt
algorithm is the fundamental tool in the theory, and improvements to
efficiency and practicality of this algorithm is the most important
problem to be faced from an applied point of view. A promising approach
to this problem can be found in the thesis of Lisle \cite{Lisle}, who
rewrites the systems of \PDES\ using first order noncommutative
operators to effect an improvement in efficiency.

The next section contains an outline of the theory of differential
Gr\"obner bases.  After establishing the notation and formulae
required, we show how to calculate the triangulation of a system of
\PDES\ using this theory. A variation of the Kolchin-Ritt algorithm is
given that appears to be much faster than the one presented in
\cite{CF,MF}, at least for the examples calculated in this paper, and
has been developed especially for use in solving the determining
equations arising from the nonclassical method. Such systems of
determining equations appear to have a natural ordering which can be
exploited to improve the efficiency of the calcualtions. We also give a
``skeletal" form of the Kolchin-Ritt algorithm which we found useful in
extracting information to reduce the complexity of the calculations. We
refer to this second algorithm as ``Direct Search'', which appears to
be new.

  In the third section of this paper, classical symmetries of the
  general nonlinear heat equation are catalogued for all those
nonlinear $f(u)$ where a non-trivial symmetry is possible. Similarly,
nonclassical symmetries are given in the  fourth section, including
some nonclassical symmetries for the linear heat equation. A catalogue
of nonclassical symmetry reduction solutions, for cubic $f$, is given
in the fifth section; the classification is in terms of the roots of
$f(u)$.  The paper concludes with some plots of several exact solutions
of (1.1) derived from nonclassical symmetry reductions.

There is much current interest in the determination of symmetry
reductions of \PDES\ which reduce the equations to \ODES. One
frequently then checks if the resulting \ODE\ is of {\it\p\ type\/}
(i.e., its solutions have no movable singularities other than poles).
It appears to be the case that whenever the \ODE\ is of \p\ type, or
can be transformed to one that is, then it can be solved explicitly,
leading to exact solutions to the original equation. Conversely, if the
resulting \ODE\ is not of \p\ type, then often one is unable to solve
it explicitly.

Since solutions of \PDES\ asymptotically tend to solutions of
lower-dimensional equations obtained by symmetry reduction, some of
these special solutions will illustrate important physical phenomena.
In particular, for reaction-diffusion equations such as (1.1), exact
solutions arising from symmetry methods can often be effectively used
to study properties such as ``blow-up'' \cite{Galaka,Galakb,GDEKS}.
Furthermore, explicit solutions (such as those found by symmetry
methods) can play an important role in the design and testing of
numerical integrators; these solutions provide an important {practical}
check on the accuracy and reliability of such integrators
(cf.\ \cite{Ames2,RAmes,Shok}).

Several exact solutions of equations of the form (\ref{nheateq}) have
been derived in the literature. Ablowitz and Zeppetella \cite{AbZ}
obtained an exact travelling wave solution of Fisher's equation
\cite{Fish} \beq u_t = u_{xx} + u(1-u) \label{fisher}\eeq by finding
the special wave speed for which the resulting \ODE\ is of \p-type.
Recently Guo and Chen \cite{GuoC} have used the Painlev\'e expansion
method \cite{NTZ,Weiss,WTC} to obtain some heteroclinic and homoclinic
solutions of (\ref{fisher}).  Kaliappan \cite{Kal} and Herrera, Minzoni
and Ondarza \cite{HMO} have derived travelling wave solutions of the
generalized Fisher's equations \bear u_t &=& u_{xx} +
u-u^k,\label{fisher1}\\ u_t &=& u_{xx} + u^p-u^{2p-1} \label{fisher2}
\eear respectively. Cariello and Tabor \cite{CTa,CTb} found an exact
solution of the real Newell-Whitehead equation \cite{NW} (or
Kolmogoroff-Petrovsky-Piscounov equation \cite{KPP}) equation \beq u_t
= u_{xx} + u(1-u^2) \label{rnw}\eeq
 using a truncated Painlev\'e expansion and verified that it derives
from a nonclassical symmetry reduction (see also \cite{Conte}).
Several authors have studied exact solutions of the Fitzhugh-Nagumo
equation \beq u_t = u_{xx} + u(1-u)(u-a) \label{fitznag}\eeq where $a$
is an arbitrary parameter, which arises in population genetics
\cite{AW,AWii} and models the  transmission of nerve impulses
\cite{Fitz,NAY}. Travelling wave solutions of the Fitzhugh-Nagumo
equation (\ref{fitznag}) have been studied by several authors
\cite{AW,AWii,FMcL,HF,Vel81}. Exact solutions of (\ref{fitznag}) have
been  obtained using various techniques including Vorob'ev \cite{Vor}
(who calls the associated symmetry a ``partial symmetry of the first
type''), by Kawahara and Tanaka \cite{KT} using Hirota's bi-linear
method \cite{Hirota}, by Hereman \cite{Hereman} using the truncated
Painlev\'e expansion method (see also \cite{CG}) and more recently by
Nucci and Clarkson \cite{NC} using the nonclassical method (see also
\cite{CM}).  Exact solutions of the Huxley equation \beq u_t = u_{xx} +
u^2(1-u) \label{huxley}\eeq have been obtained by Chen and Guo
\cite{CG}, using a truncated Painlev\'e expansion (see also
\cite{Chowd,EG}), and by Clarkson and Mansfield \cite{CM} using the
nonclassical method. Exact solutions of equation \beq u_t = u_{xx} -u^3
\label{cubiceq}\eeq have been obtained by Clarkson and Mansfield
\cite{CM} using the nonclassical method.

A number of symbolic manipulation programmes have been developed,
e.g.\ in {\sc macsyma} \cite{CHW}, {\sc maple}
\cite{CDF,Reida,Reidb,Reidc}, {\sc mathematica} \cite{Bau,Herod}, {\sc
mumath} \cite{Head} and {\sc reduce}
\cite{Ker,Nucci,Scha,Schb,Schc,Sher}, that calculate the determining
equations for classical Lie symmetries of \PDES.
  A  survey of the different packages presently available and a
discussion of their strengths and applications is given in
\cite{Here}.  The determining equations for symmetries solved in this
paper were calculated using the {\sc macsyma}  programme {\sc
symmgrp.max} \cite{CHW}.  The triangulations of these systems of \PDES\
were calculated using the {\sc maple} package diffgrob2 \cite{Mp}.  The
plots of some of the exact solutions of (1.1) derived here were drawn
using {\sc maple} and {\sc mathematica}.

\newsection{Differential Gr\"obner Bases} In this section we discuss
the triangulation of a system of \PDES\ from which solutions may be
obtained in a systematic way and give some algorithms using which a
triangulation can be determined. First we introduce some required
notation.

\subsection{Some Notation} The determining equations for symmetries of
\PDES\ can be regarded as polynomials in the infinitesimals $\xi$,
$\tau$ and $\phi$, their derivatives, and the variables $\{x,t,u\}$,
with complex coefficients.  We establish our notation and formulae
required for the statement of the algorithms in generic co-ordinates
for ease of reference.

Let $f$ be a differential polynomial; that is, a polynomial with
complex coefficients in the independent variables $\{x_1,\dots ,
x_n\}$, the functions $\{u_1,\dots, u_m\}$ and the derivative terms
$\{\mbox{D}^{\alpha}u_j\}$ where \beqn
\mbox{D}^{\alpha}u_j={\pa^{|\alpha|}u_j\over{\pa x_1^{\al_1}\dots \pa
x_n^{\al_n}}}.  \eeqn The set of all such differential polynomials
(d.p.'s) is denoted $R_{n,m}$ or\hfill\break
$\Cdiff(x_1,\dots,x_n;u_1,\dots,u_m)$.

All the concepts depend upon an ordering on the derivative terms. There
are many suitable orderings for different purposes. They need to
satisfy $\mbox{D}^{\alpha}u_j > \mbox{D}^{\beta}u_k$ implies that
$\mbox{D}^{\gamma}\mbox{D}^{\alpha}u_j >
\mbox{D}^{\gamma}\mbox{D}^{\beta}u_k$ and $\mbox{D}^{\gamma}u_j <
\mbox{D}^{\gamma}\mbox{D}^{\alpha}u_j$. We assume that $u_j > x_k$ for
all $j$ and $k$. The standard lexicographic ordering based on
$u_1<u_2<\dots <u_m$ and $\q x_1<x_2<\dots <x_n$ is given by \bearn
&&\mbox{D}^{\alpha}u_j > \mbox{D}^{\beta}u_k \\ &&\mbox{if}\ u_j>u_k,
\\ &&\mbox{else}\quad j=k \and \alpha_n>\beta_n, \\ &&\mbox{else}\quad
\al_n=\be_n,\dots,\al_{n-j}=\be_{n-j},
\al_{n-j-1}>\be_{n-j-1}\ \mbox{for\ some}\ j \ \mbox{such\ that}\ 0\le
j\le n-2.  \eearn

The {\em highest derivative} term occurring in a d.p., $f$, is denoted
$\mbox{HDT}(f)$.

The {\em highest power} of the $\mbox{HDT}(f)$ occurring in $f$ is
denoted $Hp(f)$.

The {\em highest coefficient}, $H\mbox{coeff}(f)$, is defined to be
$\mbox{coeff}(f,\,\mbox{HDT}(f)^{Hp(f)})$.

The {\em head} of $f$ is  Head$(f)=H\mbox{coeff}(f)
\mbox{HDT}(f)^{Hp(f)}$.

The {\em separant} of $f$, the highest coefficient of $\mbox{D}^\al f$
for any non-zero multi-index $\al$, is denoted $Sep(f)$.

A {\em pseudo-reduction} of a d.p.\ $f$ by $G =
\{g_1,\,g_2,\dots,g_k\}\subset R_{n,m}$ effects elimination from $f$ of
any terms that are of the form $\mbox{D}^\al \mbox{HDT}(g_i)$, for some
$i\in \{1,\dots,k\}$ and $\al\in  \N^n$.

Let a derivative term $DT$ occur in $f$ to some power $p$. Suppose
there exists an $\al\in \N^n$ such that $\mbox{D}^\al \mbox{HDT}(g) =
DT$ for some $g \in G$.  If $\al = 0$ assume further that $p \ge
Hp(g)$.  A pseudo-reduction to $f'$ of $f$ by $g$ is given by the
formulae \beqn f'=\cases{
       {\displaystyle{H\mbox{coeff}(\mbox{D}^\al g) f -
       \mbox{coeff}(f,\,DT^p) DT^{p-1}
	\mbox{D}^\al g}\over{\displaystyle Z^{\phantom2}}},     &   if
	$\al \ne 0$,\hfill\cr {\displaystyle{H\mbox{coeff}(g) f -
       \mbox{coeff}(f,\,DT^p) \mbox{HDT}(g)^{(p-Hp(g))}
       g}\over{\displaystyle Z^{\phantom2}}},  & if $\al = 0$\hfill\cr
}\eeqn where $Z$ is the great\-est com\-mon divisor of
$H\mbox{coeff}(\mbox{D}^\al g)$ and $\mbox{coeff}(f,\,DT^p)$.

The {\em pseudo-normal} form of $f$ with respect to a set $G$ is
obtained when no further pseudo-reduction with respect to any member of
$G$ is possible, and is denoted normal$^p(f,G)$.

The {\em diff$S$polynomial} of two d.p.'s is obtained first by
cross-diff\-er\-en\-tia\-ting and then by cross-multiplying by the
highest coefficients, and subtracting.

\def\cb{\hbox} Let $f_1$ and $f_2$ be two d.p.'s with the same highest
unknown.  Let $\al_1$ and $\al_2$ be the smallest multi-indices
possible such that $\mbox{D}^{\al_1} \mbox{HDT}(f_1) = \mbox{D}^{\al_2}
\mbox{HDT}(f_2)$.

 Let $Z =
\gcd(H\cb{coeff}(\mbox{D}^{\al_1}f_1),\,H\cb{coeff}(\mbox{D}^{\al_2}f_2))$.
 If both $\al_1,\, \al_2 \ne 0$, define \beqn
\cb{diff}S\cb{poly}(f_1,f_2)  =
{\displaystyle{H\cb{coeff}(\mbox{D}^{\al_1}f_1) \mbox{D}^{\al_2}f_2-
H\cb{coeff}(\mbox{D}^{\al_2}f_2) \mbox{D}^{\al_1}f_1}\over{Z}}.  \eeqn
	If $\al_1 = 0$ and $\al_2 \ne 0$, then \beqn
	\cb{diff}S\cb{poly}(f_1,f_2)  = {\displaystyle{H\cb{coeff}(f_1)
	\mbox{HDT}(f_1)^{(Hp(f_1)-1)} \mbox{D}^{\al_2} f_2 -
H\cb{coeff}(\mbox{D}^{\al_2}f_2) f_1}\over{Z}} \eeqn and similarly if
$\al_1 \ne 0$ and $\al_2 = 0$.

If $\al_1 = \al_2 = 0$ so that $\mbox{HDT}(f_1) = \mbox{HDT}(f_2)$, or
if $f_1$ and $f_2$ have different highest unknowns, then the
differential $S$ polynomial is defined to be \beqn
\cb{diff}S\cb{poly}(f_1,\,f_2) =  {\displaystyle{\cb{Head}(f_2) f_1-
\cb{Head}(f_1) f_2}\over{\gcd(\cb{Head}(f_1),\,\cb{Head}(f_2))}}\,.
\eeqn For linear equations with different highest unknowns, we take
their diffSpolynomial to be zero.

For $G\subset R_{n,m}$ let $S(G)$ be the multiplicative set in
$R_{n,m}$ generated by the set of factors of all the highest
coefficients and separants of the $g$ in $G$. We assume that
$\{1,-1\}\in S(G)$.

\subsection{The Triangulation of a System of PDEs} Given a system of
\PDES, $\Sigma\subset R_{n,m}$, we define the ideal generated by
$\Sigma$ to be all those equations that can be obtained from the
elements of $\Sigma$ by differentiating and adding, and multiplying by
arbitrary elements of $R_{n,m}$.
  In other words, \beqn I(\Sigma)=\Bigl\{ \sum_{\al,i} g_{\al,i}
\mbox{D}^\al f_i : f_i\in \Sigma,\, g_{\al,i}\in R_{n,m},\, \al\in
{\N}^n\Bigr\}.\eeqn  If the system is linear, we allow multiplication
by polynomials in the variables $x_1,\dots,x_n$ over ${\C}$ only.

A {\em differential Gr\"obner basis} of $I(\Sigma)$ is defined to be a
set of generators $G$ of $I(\Sigma)$ such that every element of
$I(\Sigma)$ pseudo-reduces to zero with respect to $G$.

Differential Gr\"obner Bases (\DGBS) of ideals of differential
polynomials  have many useful properties and can be used to solve, in
theory at least, a great many problems.  These include being able to
find all elimination ideals, integrability conditions and all
compatibility conditions of a system of nonlinear \PDES\ \cite{MF}.

A \DGB\ of a system of \PDES\ depends on the ordering of derivative
terms.  Given a \DGB\ for a system $\Sigma$ of nonlinear \PDES, we have
the following theorem \cite{MF}:

\noindent{THEOREM}: {\bf Elimination Ideals}.  {\sl If $\Sigma$ is a
\DGB\ for $I(\Sigma)$ in a lexicographic ordering with $u_1<u_2<\dots
<u_m$ and $x_1<x_2<\dots < x_n$, then} $\Sigma_{p,q}:=\Sigma \cap
\Cdiff(x_1,\dots,x_p,u_1,\dots,u_q)$ {\sl generates, up to} $S(\Sigma)$
$I_{p,q}= I(\Sigma) \cap \Cdiff(x_1,\dots,x_p,u_1,\dots,u_q)$. {\sl
That is, for every} $f\in I_{p,q}$ {\sl there exist an} $s\in
S(\Sigma)$ {\sl such that} $sf\in\Sigma_{p,q}$ {\sl that is,  $sf$ is
in the ideal generated by $\Sigma_{p,q}$}.\enskip $\square$

In other words, every condition involving only the first $q$ unknowns
with derivatives with respect to the first $p$ variables, obtainable
from the  original set of \PDES\ by differentiating, adding,
multiplying by like d.p.'s and so on, can be ``read off" from a
\DGB\ of the system in a lexicographic order, up to $S(\Sigma)$, which
represents the ``margin of error'' in the theory.

For systems $\Sigma$ that are linear in their highest derivative terms,
the set $S(\Sigma)$ is trivial.  The difficulties posed by non-trivial
$S(\Sigma)$ are caused by the use of pseudo-reduction in which one is
allowed to multiply the expression by non-trivial coefficients.
However replacing pseudo-reduction by reduction, in which only
multiplication by constants or polynomials in the independent variables
is allowed, leads to an algorithm that will not terminate on general
nonlinear systems. Even if terminated artificially, algorithms using
reduction have two additionally problems. First, the expression swell
renders the method impractical even for small problems, while second,
there is currently no method for deciding at which level of
prolongation of the system to terminate.

The sets $I(\Sigma) \cap \Cdiff(x_1,\dots,x_p,u_1,\dots,u_q)$ are
called the elimination ideals of $I(\Sigma)$, while the \DGB\ listed in
decreasing order, with respect to the standard lexigraphic ordering
described above, is called the ``triangulation" of the system
\cite{Reida}.

\subsection{The Algorithms} The basic algorithm used to calculate a
\DGB\ is the Kolchin-Ritt algorithm \cite{CF,Mans,MF,Reida,Reidb}.  In
this algorithm, pairs of d.p.'s are cross-differentiated and
cross-multiplied to make the leading derivative terms in each d.p.
cancel.  The result is then pseudo-reduced with respect to all other
d.p.'s to obtain a new d.p.  This is done systematically until the
result of cross-diff\-er\-en\-tia\-ting all pairs of d.p.'s given or
obtained pseudo-reduce to zero.

In practice, there are three difficulties in calculating a \DGB.  The
first is the problem of the build-up of differential coefficients
obtained by repeated cross-multi\-plica\-tion and pseudo-reduction.  If
the coefficient of a highest derivative term of any d.p.\ contains
 a derivative term, then Kolchin-Ritt no longer necessarily suffices to
calculate a \DGB. Sufficient conditions may be found in \cite{MF}, but
there remains the difficulty that one of the sufficient conditions
given in \cite{MF}\ is not algorithmic.  This condition is that none of
the highest coefficients or separants of the equations comprising the
system lie in the ideal they generate, that is, $S(\Sigma)\cap
I(\Sigma) = \emptyset$ (cf., \S2.2). In addition, if this condition
fails then pseudo-reduction can lead to spurious zeroes in the
calculation of the algorithms, resulting in loss of information.  It is
well known that the complexity of the Buchberger algorithm for
computing the Gr\"obner basis of a polynomial ideal is doubly
exponential \cite{MM}.  Since the Kolchin-Ritt algorithm is modelled on
the Buchberger algorithm, we have that the complexity of the
Kolchin-Ritt algorithm is at least that of Buchberger's.  However the
main problem is that of ``expression swell" in which the length of the
expressions calculated exceed the memory limits of the available
computer.

In spite of these problems, a \DGB\ can often be obtained by a clever
choice of the order in which pairs are cross-differentiated.  We
present here a version of the Kolchin-Ritt algorithm which for some
systems is much faster than that given in \cite{CF,MF}.  This version
of the Kolchin-Ritt algorithm sorts the system  after each new
condition is found
 so that we are essentially finding all the lowest conditions (relative
to the order) first.  This seems to work well with those systems that
are already partially sorted with respect to some lexicographical
ordering on the dependent and independent variables, such as those in
this paper.

The output statement of the algorithm shows the limits of the
Kolchin-Ritt algorithm when one begins with, or obtains en route, an
equation where the coefficient of the highest derivative contains
differential terms. The proof of the output statement can be found in
\cite{MF}.  It shows that when the highest coefficient of a
d.p.\ vanishes, for example when considering additional constraints or
conditions, then one must re-do the calculation taking the additional
conditions into account from the beginning.  In particular, when an
equation obtained factorises, and one wishes to consider a factor that
does not include the highest derivative term for that equation, then
one must re-start the algorithm including the desired factor at the
outset.

We make some remarks about factors. It should be mentioned that if a
product of functions is zero, say $f_1f_2=0$, then in general one can
only conclude that one of $f_1$, $f_2$ must be zero if the $f_i$ are
analytic.  However, if one obtains, for example, a cubic in $u_x$, then
the continuity of $u_x$ ensures that $u_x$ must be one of the roots of
the polynomial.

\smallskip {{\noindent{\bf Algorithm Kolchin-Ritt (sorting)}
\begin{tabbing} {OUTPUT}: \= $\Sigma'\supset\Sigma$ such that \kill
{INPUT}: \> $\Sigma$ a system of d.p.'s, a termordering\\ {OUTPUT}: \>
$\Sigma'\supset\Sigma$ such that \\
 \> $S(\Sigma')\cap I(\Sigma')=\emptyset\Longrightarrow\forall f\in
 I(\Sigma')$\\ \> $\exists\ s\in S(\Sigma')$ such that $sf$
 pseudo-reduces to $0$ with respect to $\Sigma'$.\\
 \\ pairsdone \> $=\{\}$\\ $i=1$, $j=2$ \\ sort $\Sigma$ into
 increasing order\\ $\Sigma'=\Sigma$\\ while $i\ne |\Sigma'|$ do\\ \>
 let $f_k$ be the $k^{\rm th}$ element of $\Sigma'$ for $0\le k\le
 |\Sigma'|$\\ \> if not \= $\{f_i,f_j\}\in$ pairsdone then\\ \> \>
 pairsdone=pairsdone $\cup\ \{\{f_i,f_j\}\}$\\ \> \>
 $h=$diff$S$poly$(f_i,f_j)$\\ \> \> $h=$normal$^p(h,\Sigma')$ \\ \> \>
 if $h\ne 0$, \= then $\Sigma'=\Sigma' \cup \{h\}$\\ \> \> \> sort
 $\Sigma'$ into increasing order\\ \> \> \> $i=1$, $j=2$\\ \> \> else
 if $j=|\Sigma'|$ then $j=i+2$, $i=i+1$\\ \> \> else $j=j+1$\\ end
\end{tabbing}}}

We give now a related algorithm, the ``direct search" algorithm.  The
``direct search" algorithm recursively looks for conditions that are
lower in the order than those given.  Its output is a ``skeleton" of
that of Kolchin-Ritt, but often provides useful information that allows
the triangulation of the system to proceed faster.

\smallskip {{\noindent{\bf Algorithm Direct Search} \begin{tabbing}
{OUTPUT}: \=  $G \supset F$\kill {INPUT}:  \>  $F$ a set of d.p.'s, a
termordering\\ {OUTPUT}: \>  $G \supset F$\\
 \\ $G=F$\\ sort $F$ into increasing order\\ let $f_k$ be the $k^{\rm
 th}$ element of $F$ for $0\le k\le |F|$\\
  $i=|F|-1$\\ $h=f_{|F|}$\\ while $i\ne 0$ do\\ \>  while \= $h\ne 0$
 do\\ \> \> $G=G\cup\,\{h\}$\\ \> \> $k=h$\\ \> \>
 $h=$normal$^p$(diff$S$poly($h,f_i),G)$\\ \> $h=\min\{k,f_i\}$\\ \>
 $i=i-1$\\ end \end{tabbing}}}

The direct search algorithm should be regarded as a ``template''
algorithm only. It is a natural generalization of the well-known
process of setting coefficients to zero in the situation where one has
an expression that is polynomial in $\{z_z,z_2,\ldots,z_m\}$ say, with
coefficients that are independent of the $z_i$.

\newsection{Classical Symmetries} The determining equations for the
classical symmetries of the nonlinear heat equation (1.1) are \bearn &
& f_1:\q \tau_u=0, \\ & & f_2:\q \tau_x=0, \\ & & f_3:\q
\tau_t-2\xi_x=0, \\ & & f_4:\q \xi_u=0, \\ & & f_5:\q \phi_{uu}=0, \\ &
& f_6:\q 2\phi_{xu}+\xi_t-\xi_{xx}=0, \\ & & f_7:\q
\phi_t-\phi_{xx}+\phi_uf-\phi \fu  -2f\xi_x=0.  \eearn These equations
were calculated using a {\sc macsyma} programme {\sc symmgrp.max}
\cite{CHW}.  The {\sc spde} package in {\sc reduce}
\cite{Scha}\ appears unable to handle equations such as (1.1) which
contain arbitrary functions.

Assume the lexicographic ordering given by $\phi>\tau>\xi>f$ and
$x>t>u$.
{}From $f_2$ and $f_3$ we immediately obtain
\beqn f_8:\q\xi_{xx}=0.\eeqn Calculating diff$S$poly$(f_6,f_7)$ we
obtain \beqn f_9:\q -2\phi_{ut}+2\phi \fuu +4\fu \xi_x-\xi_{xt}=0.\eeqn

Performing the ``direct search" algorithm on $f_5$ and $f_9$ and
reducing the results with respect to $\{\xi_u, \xi_{xx}, \phi_{uu}\}$
yields \bearn & & k_1:\q \phi_u\fuu +\phi \fuuu +2\fuu \xi_x=0, \\ & &
k_2:\q \left\{-\fuu \fuuuu +2\left(\fuuu\right)^2\right\}\phi+2\fuu
\fuuu \xi_x=0, \\ & &
k_3:\q-\xi_x\left(\fuu\right)^2\left\{\fuuuu\left(\fuuu\right)^2-2\fuu
\left(\fuuuu\right)^2 +\fuu \fuuu \fuuuuu\right\}=0.  \eearn

Performing a ``direct search" algorithm on $f_6$ and $f_9$ and reducing
the results with respect to $\{\xi_u, \xi_{xx}, \phi_{uu}\}$ yields
\bearn & & j_1:\q \phi_{xu}\fuu +\phi_x\fuuu =0, \\ & & j_2:\q
\xi_{tt}+2\phi_x\fuu =0, \\ & & j_3:\q \fuuu
\xi_{tt}+\left(\fuu\right)^2\xi_t=0.  \eearn

Since $\xi_u=0$ we must have, differentiating $j_3$ with respect to $u$
and cross-multiplying, that \beqn j_4:\q
2\xi_t\xi_{tt}\left\{\left(\fuuu\right)^2-\fuu \fuuuu \right\}\fuu
=0.\eeqn

We now consider the possibilities for $f(u)$; although it is possible
that other ``direct searches" will find other conditions, we can
simplify the calculation of other conditions by putting $\xi_x=0$ for
example.

We solve those factors in $k_3$ and $j_4$ involving derivatives of
$f(u)$.  Setting $\displaystyle H=\fuu$ in $k_3$ we obtain (assuming
that $H\ne 0$ and $\xi_x\ne 0$) \beqn
H_{uu}(H_{u})^2-2H(H_{uu})^2+HH_{u}H_{uuu}=0.\eeqn Dividing by
$HH_{u}H_{uu}$ we have \beqn{H_{u} \over H}-2{H_{uu}\over
H_{u}}+{H_{uuu}\over H_{uu}}=0.\eeqn This integrates easily; the
possibilities from $k_3$ are \[\begin{array}{ll} \mbox{(i)}\quad &
f(u)=au^3+bu^2+cu+d, \\ \mbox{(ii)}\quad &
f(u)=\left(au+b\right)^{n+2}+cu+d,\enskip\mbox{for}\ n\not=0,1,2,3, \\
\mbox{(iii)}\quad & f(u)=a\e^{bu}+cu+d, \\ \mbox{(iv)}\quad &
f(u)=[\ln(au+b)]/a+cu+d, \\ \mbox{(v)}\quad &
f(u)=\left[(au+b)\ln(au+b)-(au+b)\right]/a^2+cu+d, \enskip\mbox{or}, \\
\mbox{(vi)}\quad & \xi_x=0 \end{array}\] \noindent where $a$, $b$, $c$
and $d$ are arbitrary constants.

Next we look at the equation $j_4$.  In this case we obtain
\[\begin{array}{ll} \mbox{(i)}\quad & f(u)=au^2+bu+c,\\
\mbox{(ii)}\quad &
f(u)=\left[(au+b)\ln(au+b)-(au+b)\right]/a^2+cu+d,\enskip \mbox{or}, \\
\mbox{(iii)}\quad & \xi_t=0.  \end{array}\]

We shall not consider $f(u)$ linear for the classical case as the
symmetries in that case are already well known \cite{BCa}. Hence for
those solutions $f(u)$ not of the form $u\ln(u) + bu+c$ or $au^2 +
bu+c$, with $b$ and $c$ constants, we must have that $\xi_t=0$.

If $\xi_x=0$, from the conditions $k_1$ and $k_2$ one can conclude that
if $f(u)$ is not of the form $u\ln(u) + bu+c$ or $au^2 + bu+c$, then
only the trivial solution, $\xi=\xi_0$, $\tau=\tau_0$, $\phi=0$ with
$\xi_0$ and $\tau_0$ constants, is possible. This solution exists for
all choices of $f(u)$ and corresponds to the travelling wave reduction
\beqn u(x,t) = w(z),\qquad z=x-ct \eeqn where $c$ is an arbitrary
constant and $w(z)$ satisfies \beqn w'' + cw' + f(w)=0.\eeqn

The cases where $f(u)$ is of the form $u\ln(u) + bu+c$ or $au^2 + bu+c$
require further attention since then the equation $k_2$ does not
provide a condition for $\phi$.  For the other possibilities for
$f(u)$, we must re-calculate the triangulation to obtain further
conditions on $\xi$.  For the remainder of this section, $\cc1$,
$\cc2$, $\cc3$ and $\cc4$ are constants.

\noindent{\bf Case 3.1:}\quad $f(u)=u^2+bu+c$.  Here the triangulation
of the system is (in addition to $f_1$, $f_2$, $f_3$, $f_4$):  \beqn
4\phi+8u\xi_x+4b\xi_x+3\xi_{xt}=0,\qquad \xi_t=0,\qquad
\xi_{xx}=0,\qquad \xi_x(b^2-4c)=0. \eeqn Thus unless $f(u)$ is a
perfect square, the only solution to the system is the trivial one. If
$f(u)=u^2$ then \beq \xi=\cc1 x+\cc2,\qquad\tau=2\cc1
t+\cc3,\qquad\phi=-2\cc1 u. \label{sr:311} \eeq Hence for $\cc1\not=0$
(set $\cc1=1$) we obtain the symmetry reduction \beq u(x,t) =
{(t+\tfr12\cc3)^{-1}w(z)},\qquad z={(x+ \cc2)/(t+\cc3/2)^{1/2}}
\label{sr:312} \eeq where $w(z)$ satisfies \beq w'' + \tfr12 zw' + w +
w^2=0. \label{sr:313} \eeq It is easily shown that this equation is not
of \p-type.

\noindent{\bf Case 3.2:}\quad $f(u)=u\ln(u)+bu+c$.  In the case $c=0$,
the triangulation of the system (in addition to $f_1$, $f_2$, $f_3$,
$f_4$) is given by \beqn u\phi_u-\phi=0,\qquad \phi_t-\phi=0,\qquad
2\phi_x+u\xi_t=0,\qquad \xi_x=0,\qquad \xi_{tt}-\xi_t=0 \eeqn which
yields \beq \xi=\cc1 \e^t+\cc2,\qquad\tau=\cc3,\qquad
\phi=u\left(-{\tfr12}\cc1 x+\cc4\right)\e^t.  \label{sr:321} \eeq Hence
for $\cc3\not=0$ (set $\cc3=1$) we obtain the symmetry reduction \beq
u(x,t) = w(z)\exp\left\{-\tfr12\cc1x\e^t + \tfr14 \cc1^2\e^{2t} +
\left(\tfr12\cc1\cc2+\cc4\right)\e^t\right\},\qquad z=x-\cc1\e^t-\cc2 t
\label{sr:322} \eeq where $w(z)$ satisfies \beq w'' + \cc2 w' + bw +
w\ln w=0.  \label{sr:323} \eeq It is easily shown that this equation is
not of \p-type.

If $\cc3=0$ (set $\cc1=1$ and $\cc2=0$), we obtain the symmetry
reduction \beq u(x,t) = v(s)\exp\left\{-\tfr14\cc1 x^2+\cc4
x\over\cc1+\cc2\e^{-t}\right\}\qquad s = \e^t \label{sr:324} \eeq where
$v(s)$ satisfies \beq s{\d v\over\d s}=v\ln(v) + \left(b + \cc4^2
-\tfr12\right)v \label{sr:325} \eeq which has general solution \beq
v(s) = \exp\left\{ks - b - \cc4^2 + \tfr12\right\} \label{sr:325s} \eeq
with $k$ an arbitrary constant. Hence we obtain the exact solution \beq
u(x,t) = \exp\left\{k\e^{t}-\tfr14(x-2\cc4)^2 - b + \tfr12\right\}.
\eeq

The solutions for the other cases of $f(u)$ are calculated the same way
and are listed below.

\noindent{\bf Case 3.3:}\quad $f(u)=u^3+au^2+bu+c$.  Unless $ab-9c=0$
and $a^2-3b=0$ there is only the trivial solution.  The two algebraic
conditions together imply that $f(u)$ is a perfect cube.  Setting
$f(u)=u^3$ yields \beq \xi=\cc1 x+\cc2,\qquad\tau=2\cc1
t+\cc3,\qquad\phi=-\cc1 u. \label{sr:331} \eeq Hence for $\cc1\not=0$
(set $\cc1=1$) we obtain the symmetry reduction \beq u(x,t) =
{(t+\cc3/2)^{-1/2}w(z)},\qquad z={(x+ \cc2)/(t+\cc3/2)^{1/2}}
\label{sr:332} \eeq where $w(z)$ satisfies \beq w'' + \tfr12 zw' +
\tfr12w + w^3=0. \label{sr:333} \eeq It is easily shown that this
equation is not of \p-type.

\noindent{\bf Case 3.4:}\quad $f(u)=u^n +bu+c$, $n\not=0,1,2,3$. Unless
$b=c=0$ we obtain only the trivial solution. Otherwise for  $f(u)=u^n$
\beq \xi=\cc1 x+\cc2,\qquad\tau=2\cc1 t+\cc3,\qquad\phi=-{{2\cc1
u}/(n-1)}. \label{sr:341} \eeq Hence for $\cc1\not=0$ (set $\cc1=1$) we
obtain the symmetry reduction \beq u(x,t) =
{(t+\cc3/2)^{-1/(n-1)}w(z)},\qquad z={(x+ \cc2)/(t+\cc3/2)^{1/2}}
\label{sr:342} \eeq where $w(z)$ satisfies \beq w'' + \tfr12 zw' +
\frac{w}{n-1} + w^n =0. \label{sr:343} \eeq It is easily shown that
this equation is not of \p-type.

\noindent{\bf Case 3.5:}\quad $f(u)=\e^{au}+bu+c$, $a\ne 0$.  Unless
$b=c=0$ there is only the trivial solution.  If $f(u)=\e^{u}$ then \beq
\xi=\cc1 x+\cc2,\qquad\tau=2\cc1 t+\cc3,\qquad \phi=-2\cc1.
\label{sr:351} \eeq Hence for $\cc1\not=0$ (set $\cc1=1$) we obtain the
symmetry reduction \beq u(x,t) = w(z) - \ln(t+\cc3/2),\qquad z={(x+
\cc2)/(t+\cc3/2)^{1/2}} \label{sr:352} \eeq where $w(z)$ satisfies \beq
w'' + \tfr12 zw' + 1 + \e^{w}=0. \label{sr:353}  \eeq It is easily
shown that this equation is not of \p-type.

\noindent{\bf Case 3.6:}\quad $f(u)=\ln(u)+bu+c$. In this case we
obtain only the trivial solution.

\bigskip In Table 1 we summarise the infinitesimals obtained using the
classical Lie method. For completeness, we have included the case when
$f(u)$ is linear.

 \begin{table} \caption{Infinitesimals for equation (1.1) obtained
using the classical Lie method} \vspace{3pt}
\begin{tabular}{|@{\ }c@{\ }|@{\ }c@{\ }|@{\ }c@{\ }|@{\ }l@{\ }|}
\hline $f(u)$ & $\xi$ & $\tau$ & $\quad\phi$ \\*[2mm] \hline $0$ &
$\alpha x+\beta t + \gamma xt + \delta$ & $2\alpha t+\gamma t^2 +
\kappa$ &
 $\left[\lambda - \tfr12\beta x - \left(\tfr12 t + \tfr14
 x^2\right)\gamma\right] u+ \phi_0(x,t)$ \\*[2mm] $1$ & $\alpha x+\beta
t + \gamma xt + \delta$ & $2\alpha t+\gamma t^2 + \kappa$ &
$\left[\lambda - \tfr12\beta x - \left(\tfr12 t + \tfr14
x^2\right)\gamma\right] u $ \\*[2mm]  & & & $\quad + (2\alpha -
\lambda)t + \tfr12\beta xt + \left(\tfr14x^2t+\tfr32 t^2\right)\gamma$
\\*[2mm]  & & & $\quad + \phi_0(x,t)$ \\*[2mm] $u$ & $\alpha x+\beta t
+ \gamma xt + \delta$ & $2\alpha t+\gamma t^2 + \kappa$ &
 $\left[\lambda + 2\alpha t - \tfr12\beta x + \left(t^2 -\tfr12
 t-\tfr14 x^2\right)\gamma\right] u$ \\*[2mm]  & & & $\quad +
\psi_0(x,t)$ \\*[2mm] $u^n\enskip (n\not= 0,1)$ & $\alpha x+\beta$ &
$2\alpha t+\gamma$ & $\displaystyle -{2\alpha u\over n-1}$ \\*[2mm]
$\e^{u}$ & $\alpha x+\beta$ & $2\alpha t+\gamma$ & $-2\alpha$ \\*[2mm]
$u\ln(u)+bu$ & $\alpha \e^t+\beta$ & $\gamma$ & $u\left(-{\tfr12}\alpha
x+\delta\right)\e^t$ \\*[3mm] arbitrary & $\alpha$ & $\beta$ & $0$
\\*[2mm] \hline \end{tabular}

\vspace{12pt} \noindent where $\phi_0$ satisfies $\phi_{0,t} =
\phi_{0,xx}$ and $\psi_0$ satisfies $\psi_{0,t} = \psi_{0,xx}+\psi_0$.
\end{table}

\newsection{Nonclassical Symmetries} There are two sorts of
nonclassical symmetries, those where the infinitesimal $\tau$ is
non-zero, and those where it is zero.  In the first case, we can assume
without loss of generality that $\tau\equiv 1$, while in the second
case we can assume that $\xi\equiv 1$.

In this section, we solve the determining equations for the
nonclassical infinitesimals, and obtain the reductions.  Solutions are
catalogued in \S5.

\subsection{Nonclassical Symmetries, $\tau\equiv 1$} The determining
equations for the nonclassical symmetries, given $\tau\equiv 1$,  are
\bearn & & f_1:\qq \xi_{uu}=0, \\ & & f_2:\qq
\phi_{uu}-2\xi_{xu}+2\xi\xi_u=0, \\ & & f_3:\qq
2\phi_{xu}-2\phi\xi_u+3\xi_uf+\xi_t-\xi_{xx}+2\xi\xi_x=0 ,\\ & &
f_4:\qq \phi_t-\phi_{xx}+\phi_uf-\fu \phi+2\phi\xi_x-2\xi_xf=0. \eearn
These were calculated using the {\sc macsyma} programme {\sc
symmgrp.max} \cite{CHW}.

The result of the ``direct search" algorithm on $\{f_1,f_2,f_3\}$ in
the ordering $\phi>\xi>f$ and $x>t>u$ is the  equation \beqn
\xi_u\fuuuu =0. \eeqn Hence, unless $\xi_u=0$, $f(u)$ must be cubic in
$u$.  In the case $\xi_u=0$, another ``direct search" on the system led
to the equation \beqn -\xi_x\left(\fuu\right)^2\left\{-2\fuu
\left(\fuuuu\right)^2+\fuu \fuuu \fuuuuu +\fuuuu \left(\fuuu\right)^2
\right\}=0\eeqn which is the same condition $k_3$ for $f(u)$ obtained
in the classical case.  If $\xi_x=\xi_u=0$ then we obtain the condition
\beqn \phi\left\{2\left(\fuuu\right)^2-\fuu \fuuuu \right\}=0. \eeqn As
in the previous Section, we consider the various possibilities for
$f(u)$ separately.

\subsubsection{$\tau\equiv1$, $f(u)=u^3+bu^2+cu+d$} The direct search
algorithm  yields the equation  $\xi_u(2\xi_u^2+9)=0$.

\noindent{\bf Case 4.1.1i}\q $2\xi_u^2+9=0$.  Here a triangulation of
the system yields the infinitesimals \beq
\xi=\tfr32\i\sqrt{2}\,(u+b/3),\qquad\phi=\tfr32(u^3+bu^2+cu+d). \eeq
and the associated invariant surface condition is \beq
\tfr32\i\sqrt{2}\,(u+b/3)u_x + u_t - \tfr32(u^3+bu^2+cu+d)=0.
\label{isc:411} \eeq By solving this we obtain exact solutions (1.1)
with $f(u)=u^3+bu^2+cu+d$ (see \S5.1 below).

Setting $\xi_u=0$ we obtain the triangulation \bearn & &
\phi+\xi_x(u+b/3)=0, \\ & & \xi_t-3\xi_{xx}+2\xi\xi_x=0, \\ & &
3\xi_{xxx}-3\xi\xi_{xx}-\xi_x\left(b^2-3c\right)=0, \\ & &
\xi_{xx}\left(-3\xi_{xx}+2\xi\xi_x\right)=0, \\ & &
\xi\xi_x^2\left(-b^2+3c\right)\left(3c-b^2-\xi\right)
\left(9c-3b^2-2\xi^2+6\xi_x\right)=0, \\ & &
\xi_x\left(-27d-2b^3+9bc\right)=0. \eearn Unless the algebraic
condition $-27d-2b^3+9bc=0$ on the coefficients of $f(u)$ is satisfied,
the only solution is the trivial one ($\xi=\xi_0$ a constant,
$\phi=0$). So assume that $\xi_x\ne 0$ and that $-27d-2b^3+9bc=0$.
There are thus several cases to consider.  Note that if both $b^2-3c=0$
and  $-27d-2b^3+9bc=0$ then $f(u)=(u+b/3)^3$.

\noindent{\bf Case 4.1.1ii}\q
$\xi_u=0$,\ $\xi_{xx}=0$,\ $f(u)=(u+b/3)^3$.  In this case we have from
the triangulation equations that $\xi_t+2\xi\xi_x=0$ and we thus have
that \beq \xi={({x+\cc1})/({2t+\cc2})},\qquad
\phi=-{({u+b/3})/({2t+\cc2})}. \eeq These are equivalent to
(\ref{sr:311}) and thus also yield the (classical) scaling reduction
(\ref{sr:312}).

\noindent{\bf Case 4.1.1iii}\q $\xi_u=0$,\ $-3\xi_{xx}+2\xi\xi_x=0$,\
$f(u)=(u+b/3)^3$. In this case we have $\xi_t=0$ and the Kolchin-Ritt
algorithm on the equations $\left\{-3\xi_{xx}+2\xi\xi_x=0,
-3\xi_{xxx}+3\xi\xi_{xx}=0\right\}$ leads to the condition
$2\xi_x(3\xi_x-\xi^2t)=0$.  Thus the  non-trivial solution in this case
is \beq \xi={-\,{3/(x+\cc1)}},\qquad\phi={-\,{3(u+b/3)/(x+\cc1)^2}}.
\eeq Hence we obtain the nonclassical symmetry reduction \beq u(x,t) =
(x+\cc1)w(z) - \tfr13b,\qquad z=\tfr12 x^2 + \cc1 x + 3t
\label{eq:411iiii}\eeq where $w(z)$ satisfies \beq w'' + w^3=0
\label{eq:411iiis}\eeq which is solvable in terms of elliptic functions
(see \S5.2 below).

\noindent{\bf Case 4.1.1iv}\q $\xi_u=0$, $27d+2b^3-9bc=0$, $b^2-3c\ne
0$.  In this case we have \beqn \xi_t=0, \qquad
 9c-3b^2-2\xi^2+6\xi_x=0 \eeqn leading to the solution \beq
\xi=3\mu\tan(\mu x+\cc1) \qquad \phi=-\mu(3u+b)\sec^2(\mu x+\cc1) \eeq
where $\mu^2=\tfr16(b^2-3c)$, and so we obtain the nonclassical
symmetry reduction \beq u(x,t) = \mu\cot(\mu x+\cc1) w(z) -
\tfr13b,\qquad z=\sin(\mu x+\cc1)\exp(-3\mu^2 t) \label{eq:411iv}\eeq
where $w(z)$ satisfies \beq z^2w'' - 2zw' + 2w + w^3=0
\label{eq:411ivs}\eeq which is solvable in terms of elliptic functions
(see \S5.2 below).

We remark that the roots of the cubic $u^3+bu^2+cu+(9c-2b^2)b/27=0$ are
$-\tfr13b$, $-\tfr13b+\sqrt{2}\,\mu$ and $-\tfr13b-\sqrt{2}\,\mu$,
where $\mu^2=\tfr16(b^2-3c)$, i.e.\ they are collinear and the
distances from the outer roots to the central one are equal. The
condition that $27d+2b^3-9bc=0$ is the necessary and sufficient
condition on $f(u)=u^3+bu^2+cu+d$ for the existence of nonclassical
symmetry reductions of (1.1) to \ODES\ solvable in terms of elliptic
functions.  Furthermore, Nucci and Clarkson \cite{NC} show that
elliptic functions solutions exist for the Fitzhugh-Nagumo equation
(\ref{fitznag}) if either $a=-1$, $a=\tfr12$ or $a=2$, which are
precisely those the three cases when the distances from the outer roots
to the central one are equal.

\subsubsection{$\tau\equiv1$, $f(u)=au+b$} A preliminary triangulation
yields the condition $\xi_u^3=0$.  Reducing the equations with respect
to $\xi_u=0$ we obtain $\phi_{uu}=0$.  Inserting $\phi=A(x,t)u+B(x,t)$
into the equations, we obtain the system \cite{BCa,HE} \bearn & &
g_1:\qquad 2A_x+\xi_t-\xi_{xx}+2\xi\xi_x=0, \\ & & g_2:\qquad
A_t-A_{xx}+2\left(A-a\right)\xi_x=0, \\ & & g_3:\qquad
B_t-B_{xx}+2\left(B-b\right)\xi_x-aB+bA=0. \eearn Note that we no
longer have an over-determined system.  For an arbitrary function
$v(x,t)$ let $\Phi(v;\xi)=v_t-v_{xx}+2v\xi_x$, and set
$\Phi(\xi;\xi)={\eta}$.  Let $A>\xi$. Then
cross-diff\-er\-en\-tia\-ting $g_1$ and $g_2$ and reducing, we obtain
\beqn 4\xi_{xx}(A-a)=\Phi({\eta};\xi).\eeqn

\noindent{\bf Case 4.1.2i}\q $\tau\equiv1$, $\xi_{xx}\ne 0$,
$f(u)=au+b$.  Pseudo-reducing $g_1$, $g_2$ with respect to
$\xi_{xx}(A-a)=\tfr14 \Phi({\eta};\xi)$, we obtain two equations $k_1$
and $k_2$ in $\xi$ whose highest derivative terms in a {\em total
degree} ordering are $\xi_{xxxxxx}$ and $\xi_{xxxxx}$. (A total degree
ordering is determined first by the degree of the terms and then by a
lexicographic ordering.)  Reducing $k_1$ with respect to $k_2$ in this
ordering yields a condition $k_3$ for $\xi$ whose highest derivative
term is $\xi_{xxxxt}$. Now $\{k_2, k_3\}$ forms a \DGB\ for
$I(\{k_1,k_2\})$ with $S(\{k_2,k_3\})=\{\pm\left(\xi_{xx}\right)^n :\,
n\in\mbox{\N}\}$.  (We have stipulated that $\xi_{xx}\ne 0$ in this
case.)  Our next step is to apply the Initial Data algorithm
\cite{Reida}, from which we have that a formal solution to the
equations for $\xi$ depends on four arbitrary functions and one
arbitrary constant.  In fact we can assign $\xi(0,t)$, $\xi_{x}(0,t)$,
$\xi_{xx}(0,t)$, $\xi_{xxx}(0,t)$ and $\xi_{xxxx}(0,0)$ and the other
coefficients for a
 formal expansion for $\xi$ will be determined.

Assuming that $\xi_{xx}(0,t)$ is a constant, and that the other
arbitrary functions and constant are zero, one can derive
 a solution to the equations for $\xi$ and $A$ (i.e.\ $g_1$ and
 $g_2$):  \beq \xi={3x^2\over{3c-x^3}},\qquad A=a+{{3x}\over{3c-x^3}}
\label{xA:46} \eeq where $c$ is a constant, simply by summing the
formal Taylor series obtained from $k_2$ and $k_3$. Solving $g_3$ for
$B$ with $\xi$ and $A$ as given in (\ref{xA:46}) yields \beqn B=
b+{3bcx\over a(3-cx^3)}.\eeqn Therefore the infinitesimals in this case
are \beq \xi={3x^2\over{3c-x^3}},\qquad \phi=(au+b)+{3x(au+b)\over
a({3c-x^3})} \label{infins:412} \eeq and so we obtain the symmetry
reduction \beq u(x,t) = xw(z)\,\e^{at} - b/a,\qquad z = \tfr16x^2 + t +
c/x\eeq where $w(z)$ satisfies \beqn w''=0.\eeqn Hence we obtain the
exact solution of (1.1) with $f(u)=au+b$ given by \beq u(x,t) =
\left\{\alpha(\tfr16x^3 + xt + c) + \beta\right\}\e^{at} - b/a.\eeq

\noindent{\bf Case 4.1.2ii}\q $\tau\equiv1$, $\xi_{xx}=0$,
$f(u)=au+b$.  In this case we obtain a classical reduction and so we
omit details.

Nonclassical symmetry reductions for the linear equation \beq u_t =
u_{xx} + au+b \eeq where $a$ and $b$ are arbitrary constants, obtained
using both \p\ and symmetry analysis of the determining equations
$\{g_1, g_2, g_3\}$  can be found in \cite{Webba,Webbb}.

\subsubsection{Nonclassical, $\tau\equiv1$, $f(u)$ neither cubic nor
linear} In this case we have \beqn -2\fuu \left(\fuuuu\right)^2+\fuu
\fuuu \fuuuuu +\fuuuu\left(\fuuu\right)^2=0\eeqn or \beqn
2\left(\fuuu\right)^2-\fuu \fuuuu =0.\eeqn The various possibilities
for $f(u)$ have already been listed in \S 3.  For all these
possibilities, we obtain the same symmetries
 as for the classical case, with the classical  $\xi\left/\tau\right.$
yielding the nonclassical $\xi$ and the classical
$\phi\left/\tau\right.$ yielding the nonclassical $\phi$.  However, for
some of the cases, this is harder to prove than for the classical
determining equations.

We give  the proof for the case $f=u^n+cu+d$ where $n\ne 0,1,2$ or $3$,
to show the ideas involved.

We have in this case that $\xi_u=0$.  The Direct Search algorithm
yields that $d=0$ else one has only the trivial solution, and that
$\phi=-2u\xi_x/(n-1)$.

Reducing the determining equations with respect to $\xi_u$ and
$\phi+2\left/(n-1)\right.u\xi_x$ yields \bearn && h_1:\q
4\xi_{xxx}-2(n-1)\xi\xi_{xx}+c(n-1)^2\xi_x=0,\\ && h_2:\q
(n-1)\xi_t-(n+3)\xi_{xx}+2(n-1)\xi\xi_x=0.  \eearn Take the ordering
$t>x$. Then cross-diff\-er\-en\-tia\-ting $h_1$ and $h_2$, throwing
away powers of $c$, $\xi_x$ and $(n-1)$ that factor out, and iterating,
we obtain successively equations with the following ``vital
statistics":  \begin{tabbing} \baselineskip=13.5pt {$h_3$\q
XX}\={XXXX}\={XX$\mbox{HDT}(h_3)^{Hp(h_3)}$}\=\kill $h_3$:\q
\>\mbox{HDT}$(h_3)^{\mbox{\scriptsize Hp}(h_3)}=\xi_{xx}^2$,
 \>\>Sep($h_3)=24n\xi_{xx}+n^2(n-1)\xi\xi_x$,\\ $h_4$:\q
\>\mbox{HDT}$(h_4)^{\mbox{\scriptsize Hp}(h_4)}=\xi_{xx}$,\\
\>\>Sep($h_4)=-12(n+1)\xi_x+(5n^2-2n^3-5)\xi^2 +12c(n^2-120n+108)$,\\
$h_5$:\q \>\mbox{HDT}$(h_5)^{\mbox{\scriptsize Hp}(h_5)}=\xi_x^3$,
\>\>Sep($h_5)=p(\xi_x,\xi)$ \end{tabbing} where $p$ is a polynomial.
Using the Implicit Function Theorem, we obtain for some function $R$,
the ansatz $ \xi_x=R(\xi) $.  Note that if any of the separants are
zero, leading to invalid reductions, we still obtain the same ansatz.
Substituting the ansatz into $h_1$, $h_2$,  one obtains \bearn &&k_1:\q
\xi_x-R(\xi)=0,\\ &&k_2:\q  (n-1)\xi_t+2(n-1)\xi
RR_{\xi}-(n+3)RR_{\xi}=0,\\ &&k_3:\q  (n-1)R^2R_{\xi\xi}+2(n-1)\xi
RR_{\xi} -4RR_{\xi}^2 -2(n-1)R^2-n(n-2)cR=0.  \eearn
Cross-differen\-tia\-ting $k_1$ and $k_2$ ($\xi>R$) and reducing yields
\beq R^2\left[(n+3)R_{\xi\xi}-2(n-1)\right]=0. \label{4.3.11}\eeq
Solving for $R$ using the second factor of (\ref{4.3.11}), and
substituting the result into $k_3$, yields a polynomial for $\xi$, that
is, $\xi$ is a constant.

Thus, since the only factors we threw away were $c$, $\xi_x$ and
$(n-1)$, it must be that for a nontrivial solution to exist, $c=0$.

Setting $c=0$ in $h_1$ and $h_2$ and performing the Kolchin-Ritt
algorithm on these two equations yields the condition \beq
(n-1)\xi_{xx}\left[(n^2-1)\xi\xi_x-12\xi_{xx}\right].\label{4.3.1}\eeq
The system $\{\xi_{xx}=0,\xi_t+2\xi\xi_x=0\}$ has for its solution
\beqn \xi={x+\cc2\over2t+\cc1},\qq \phi=-\,{2u\over(n-1)(2t+\cc1)}\eeqn
which is equivalent to the classical reduction of Case 3.3.

Putting the other factor of (\ref{4.3.1}), namely
$(n^2-1)\xi\xi_x-12\xi_{xx}$, in addition to $h_1$ and $h_2$ (with
$c=0$) as input to the Kolchin-Ritt algorithm, we obtain \bearn &&
\xi_x(n-1)(n+1)\left[12\xi_x+(n-1)(n-5)\xi\right]=0,\\ &&
\xi^4(n-1)^3(n-3)(n-5)^5=0.  \eearn Since $n\ne 1$ or 3 in this case,
there are no other nontrivial solutions.

\subsection{Nonclassical Symmetries, $\tau\equiv0$,\ $\xi\equiv 1$}
 We come now to those nonclassical symmetries where $\tau\equiv0$,
 and`` without loss of generality we can take $\xi\equiv 1$.  In this
case we have the system consisting of the equation (1.1), the invariant
surface condition (1.4) and one additional determining equation for
$\phi$:  \bearn & & f_1:\qq u_t-u_{xx}-f(u)=0, \\ & & f_2:\qq
u_x-\phi(x,t,u(x,t))=0, \\ & & f_3:\qq
\phi_t-\phi_{xx}-\phi^2\phi_{uu}-2\phi\phi_{xu}+f\phi_u-f_u\phi=0.
\eearn  The last equation was calculated using the {\sc macsyma}
package {\sc symmgrp.max} \cite{CHW}.  Cross-diff\-er\-en\-tia\-ting
$f_1$ and $f_2$ produces exactly $f_3$ so we obtain no new conditions.
So we try three ans\"atze for $\phi$, namely

\smallskip\noindent{\bf Ansatz 4.2.1}: $\phi=p_2 u^2+p_1u+p_0$ with
$f(u)=-u^3-bu^2-cu-d$.

\smallskip\noindent{\bf Ansatz 4.2.2}: $\phi=q_1(x,t)u+q_0(x,t)$.

\smallskip\noindent{\bf Ansatz 4.2.3}: $\phi=r_1(t,u)x+r_0(t,u)$.

\smallskip\noindent In the second and third ans\"atze, $f(u)$ is to be
determined. We consider these three ans\"atze in turn.  Note that in
the equation $f_3$, $u$ is to be regarded as an independent variable.

\smallskip\noindent{\bf Ansatz 4.2.1}. Reducing $f_3$ using the ansatz,
and setting coefficients of powers of $u$ to zero, we obtain from the
coefficient of $u^4$ that $p_2(2p_2^2-1)=0$. Setting
$p_2=\tfr12\sqrt{2}$ we have from the coefficients of $u^2$, $u$ and 1
that:  \bearn && g_1:\qquad 2p_1^2-\sqrt{2}\,bp_1-\sqrt{2}\,p_0+c=0,
\\ && g_2:\qquad 2p_0p_1-\sqrt{2}\,bp_0+d=0, \\ && g_3:\qquad
\sqrt{2}\,p_0^2-cp_0+dp_1=0.\nonumber \eearn The coefficient of $u^3$
is identically zero. For generic $b$, $c$ and $d$ there are three
possible solutions of $\{g_1,g_2,g_3\}$ given by \beq
p_1=-\tfr12\sqrt{2}\,(m_i+m_j),\qquad p_0=\tfr12\sqrt{2}\,m_im_j,
\qquad 1\leq i < j \leq 3, \label{cc12}  \eeq  where $m_i$ and $m_j$
are any two of the three distinct roots of the cubic \beq
u^3+bu^2+cu+d=0.\eeq Hence there are three choices for $\phi$ \beq \phi
= \tfr12\sqrt{2}\,(u-m_i)(u-m_j),\qquad 1\leq i < j \leq 3.  \eeq
Substituting this into $f_2$, solving for $u$ and requiring that $u$
also satisfies $f_1$ yields the solution \beq u_t = u_{xx}
-u^3-bu^2-cu-d= u_{xx} - (u-m_1)(u-m_2)(u-m_3),\eeq given by \beq
u(x,t) = {m_{i}\cc{i}\Theta_{i}(x,t) + m_{j}\cc{j}\Theta_{j}(x,t)\over
 \cc{i}\Theta_{i}(x,t) +\cc{j}\Theta_{j}(x,t)},\qquad 1\leq i < j \leq
 3,\label{t0sol}\eeq where \beqn \Theta_k(x,t) =
\exp\left\{\tfr12\sqrt{2}\,m_k x - m_k\left(m_1+m_2+m_3 -
\tfr32m_k\right)t\right\},\qquad 1\leq k \leq 3.\eeqn We remark that
the solution (\ref{t0sol}) is special case of the travelling wave
solution and of the exponential solution ({\ref{sol:512}) given in
\S5.1.1 below.

If $d=(9c-2b^2)b/27$, then there is the solution of $\{g_1,g_2,g_3\}$
given by $p_1=\tfr13\sqrt{2}\,b$ and $p_0=\tfr12\sqrt{2}\,(c-\tfr29
b^2)$. Thus, a similar procedure to the generic case above yields exact
solutions of \beqn u_t = u_{xx} - u^3 - bu^2 - cu -(9c-2b^2)b/27,\eeqn
given by \beqn u(x,t) =
\cases{(c-\tfr13b^2)^{1/2}\tan\left(\tfr12\sqrt{2}\,(c-\tfr13b^2)^{1/2}
x + \cc1\right)- \tfr13 b, & if $c\not=\tfr13b^2$ \cr
{\displaystyle\pm\,{\sqrt2\over x + \cc2} - \tfr13 b}, & if
$c=\tfr13b^2$ \cr}\eeqn where $\cc1$ and $\cc2$ are arbitrary
constants. We remark that $d=(9c-2b^2)b/27$ is the same relation as
that we obtained for (1.1) with $f(u)$ cubic and $\tau\equiv1$ which
yielded special reductions in terms of elliptic solutions.

If $p_1=0$ then $p_0=\tfr12\sqrt{2}\,c$ provided that $d=bc$. Hence we
obtain the exact solution of \beqn u_t = u_{xx} - u^3 - bu^2 - cu
-bc,\eeqn given by \beqn u(x,t) =
\sqrt{c}\,\tan\left[\sqrt{c}\,(\tfr12\sqrt{2}\,x-bt)\right],\eeqn which
is a special case of the travelling wave reduction.

If $p_2=0$, we then have that $p_1=p_0=0$ implying that $\phi=0$. Thus
solving the equations $f_1$ and $f_2$ yields \bearn &&
u(x,t)=g(t),\qquad\mbox{where}\qquad {\d g\over\d t}+ag^3+bg^2+cg+d=0.
\eearn

\smallskip\noindent{\bf Ansatz 4.2.2}. In this ansatz, we take \beq
\phi=q_1(x,t)u+q_0(x,t) \eeq Reducing $f_3$ using the ansatz, it can be
shown that there are three cases to consider.

\smallskip\noindent{\bf Case  4.2.2i} $f=au^2+bu+c$, $a\ne 0$. In this
case we have $\phi=0$ and solving the equations $f_1$ and $f_2$ yields
\bearn && u(x,t)=g(t),\qquad {\d g\over\d t}=ag^2+bg+c.  \eearn

\smallskip\noindent{\bf Case  4.2.2ii} $f=au+b$. In this case, it is
easily shown that $\phi=(u+b/a)q_1$, where $q_1$ satisfies $$q_{1,t} -
q_{1,xx} - 2 q_{1,x} q_1=0.$$  Combining this with the equations $f_1$
and $f_2$, we recover the well-known Cole-Hopf transformation
\cite{Cole,Hopf} relating a solution of Burgers' equation to a solution
of the linear heat equation.

\smallskip\noindent{\bf Case  4.2.2iii} $f(u) = (u + k)\ln(u + k) -(u +
k) + d u + e$. In this case, it is easily shown that  if  $e=dk$ then
$\phi=(u+k)q_1(x,t)$ where $q_1$ satisfies $$q_{1,xx}  + q_1  -
q_{1,t}  + 2 q_1 q_{1,x}=0,$$ whilst if $e\not=dk$ then $\phi=0$.  The
situation is analogous to that in Case 4.2.2ii above.

It appears that B\"acklund transformations arise in considering
nonclassical symmetries in the $\tau=0$ case. Further  interesting
examples have been obtained by Nucci \cite{Nuccix}.

\smallskip\noindent{\bf Ansatz 4.2.3} $\phi=r_1(t,u)x+r_0(t,u)$.
Reducing $f_3$ using the ansatz, it can be shown that there are three
cases to consider:  (i), $f(u)=au^2+bu+c$, $a\ne 0$,  (ii),
$f(u)=au+b$, $a\ne 0$, or  (iii), $f(u)=u\ln(u)-u+cu+d$.

\smallskip\noindent{\bf Case 4.2.3i} $f(u)=au^2+bu+c$, $a\ne 0$. In
this case, $\phi=0$.

\smallskip\noindent{\bf Case 4.2.3ii} $f(u)=au+b$, $a\ne 0$. In this
case we obtain the exact solution of \beq u_t = u_{xx} + au+b,\eeq
given by \beq u(x,t) = {\cc4\over(t+\cc2)^{1/2}}\exp\left\{ at -
{(x+\cc1)^2\over4(t+\cc2)}\right\} - {b+\cc3\,\e^{at}\over a}, \eeq
where $\cc1,\cc2,\cc3$ and $\cc4$ are arbitrary constants.

\smallskip\noindent{\bf Case 4.2.3iii} $f(u)=(u+k)\ln(u+k)+c(u+k)$. In
this case we obtain the exact solution of \beq u_t = u_{xx} +
(u+k)\ln(u+k)+c(u+k),\eeq given by \beq u(x,t) =\exp\left\{\cc2\e^t -
\tfr14(x+\cc1)^2 - c+\tfr12\right\}-k \eeq where $\cc1$ and $\cc2$ are
arbitrary constants.

\def\sech{\mathop{\mbox{sech}}\nolimits}
 \begin{table} \caption{Infinitesimals for equation (1.1) obtained
using the nonclassical method} \vspace{3pt}
\begin{tabular}{|@{\ }c@{\ }|@{\ }c@{\ }|@{\ }c@{\ }|@{\ }l@{\ }|}
\hline $f(u)$ & $\xi$ & $\tau$ & $\quad\phi$ \\*[2mm] \hline
$u^3+bu^2+cu+d$ & $\tfr32\i\sqrt{2}\,(u+b/3)$ & $1$ &
 $\tfr32(u^3+bu^2+cu+d)$ \\*[2mm] $-u^3-bu^2-cu-d$ &
$\tfr32\sqrt{2}\,(u+b/3)$ & $1$ &
 $-\tfr32(u^3+bu^2+cu+d)$ \\*[2mm] $u^3$ & $\displaystyle -\,{3\over
x+\cc1}$ & $1$ &
 $\displaystyle -\,{3(u+b/3)\over(x+\cc1)^2}$ \\*[3mm] $-u^3$ &
$\displaystyle -{3\over x+\cc1}$ & $1$ &
 $\displaystyle {3(u+b/3)\over(x+\cc1)^2}$ \\*[3mm]
$u^3+bu^2+cu+b(9c-2b^2)/27$ & $3\mu\tan(\mu x+\cc3)$ & $1$ &
 $-\mu(3u+b)\sec^2(\mu x+\cc3)$ \\*[2mm] $-u^3-bu^2-cu-b(9c-2b^2)/27$ &
$3\mu\tanh(\mu x+\cc4)$ & $1$ &
 $\mu(3u+b)\sech^2(\mu x+\cc4)$ \\*[2mm] $ au+b $ & $\displaystyle
{3x^2\over 3\cc5 - x^3}$ & $1$ & $\displaystyle au+b + {3x(au+b)\over
a(3\cc5 - x^3)}$ \\*[3mm] $-(u-m_1)(u-m_2)(u-m_3)$ & $1$ & $0$ &
$\tfr12\sqrt{2}\,(u-m_1)(u-m_2)$ \\*[2mm] $-(u-m_1)(u-m_2)(u-m_3)$ &
$1$ & $0$ & $\tfr12\sqrt{2}\,(u-m_2)(u-m_3)$ \\*[2mm]
$-(u-m_1)(u-m_2)(u-m_3)$ & $1$ & $0$ & $\tfr12\sqrt{2}\,(u-m_3)(u-m_1)$
\\*[2mm] $-u^3-bu^2-cu-b(9c-2b^2)/27$ & $1$ & $0$ &
 $\tfr12\sqrt{2}\left[(u+b/3)^2 + (c-\tfr13b^2)\right]$ \\*[2mm]
$-u^3-bu^2-cu-bc$ & $1$ & $0$ &
 $\tfr12\sqrt{2}\,(u^2 + c)$ \\*[2mm] $ (u+k)\ln(u+k) + c(u+k) $ & $1$
& $0$ & $-\tfr12(x+\cc6)(au+b)$ \\*[2mm] $ au+b $ & $1$ & $0$ &
$\displaystyle -\,{(x+\cc8)(au+b+\cc9\e^{at})\over2(t+\cc{10})}$
\\*[3mm] \hline
 \end{tabular}

\vspace{12pt} \noindent where $\cc1,\cc2,\ldots,\cc{10}$ are arbitrary
constants.  \end{table}

\newsection{Exact solutions of $u_t = u_{xx} +u^3+bu^2+cu+d$} In this
section we obtain exact solutions of the \PDES\ \beq u_t = u_{xx}
+u^3+bu^2+cu+d \label{pde:511} \eeq and \beq u_t = u_{xx} -
(u^3+bu^2+cu+d) \label{pde:512} \eeq using the infinitesimals derived
in \S4.1.1 above. There are two types of solutions:  those expressible
in terms of exponentials arising from Case 4.1.1i (\S5.1)  and those
expressible in terms of elliptic functions arising from Cases 4.1.1iii
and 4.1.1iv (\S5.2).  It is interesting to note that the form of the
exponential-type solution is closely related to the roots of the cubic
\beq u^3+bu^2+cu+d=0 \label{cubic:u} \eeq arising in the \PDE\ we are
considering. Furthermore the conditions when
(\ref{pde:511},\ref{pde:512}) possess solutions expressible in terms of
elliptic functions are determined by the roots of (\ref{cubic:u}).
Recall that nonclassical symmetry reductions of (5.1), (5.2) are
obtainable in the  case when the roots of $f(u)$ are arbitrary (Case
4.1.1i),
 and when the roots of $f(u)$ are collinear and the distances from the
outer roots to the middle root are equal (Case 4.1.1iv).  These
reductions lead to exponential and elliptic type solutions
respectively.  The second case is determined by the equivalent
algebraic condition $27d+2b^3-9bc=0$ on the coefficients of
$f(u)=u^3+bu^2+cu+d$.  We note that a solution of (\ref{pde:512}) can
be obtained from one of (\ref{pde:511}) by letting $x\to\pm\i x$ and
$t\to -t$, and vice-versa.

\subsection{Exponential-type solutions} In Case 4.1.1i above, we
obtained the infinitesimals for (\ref{pde:511}) given by \beq
\xi=\tfr32\i\sqrt{2}\,(u+b/3),\qquad\phi=\tfr32(u^3+bu^2+cu+d)
\label{inf:511} \eeq with associated invariant surface condition \beq
\tfr32\i\sqrt{2}\,(u+b/3)u_x + u_t - \tfr32(u^3+bu^2+cu+d)=0.
\label{isc:511} \eeq Following Nucci and Clarkson \cite{NC}, this is
solvable as follow.  Using (\ref{isc:511}) to eliminate $u_t$ in
(\ref{pde:511})\ yields \beqn u_{xx} + \tfr32\i\sqrt{2}\,(u+b/3)u_x -
\tfr12(u^3+bu^2+cu+d)=0 \eeqn which can be linearized by the
transformation $u=-\sqrt{2}\,\i\,\Phi_x/\Phi$ yielding  \beq
2\sqrt{2}\,\Phi_{xxx} + 2\i b\Phi_{xx} - \sqrt{2}\, c\Phi_x - \i d
\Phi=0. \label{eq:phi}\eeq Suppose that the roots of the cubic
(\ref{cubic:u}) are $m_1$, $m_2$ and $m_3$, then the roots of \beq
2\sqrt{2}\,p^3 + 2\i bp^2 - \sqrt{2}\, cp - \i d =0\label{cubic:p} \eeq
are $\tfr12\sqrt{2}\,\i\,m_1$, $\tfr12\sqrt{2}\,\i\,m_2$ and
$\tfr12\sqrt{2}\,\i\,m_3$.

The analogous infinitesimals for (\ref{pde:512}) are \beq
\xi=\tfr32\sqrt{2}\,(u+b/3),\qquad\phi=-\tfr32(u^3+bu^2+cu+d)
\label{inf:512} \eeq with associated invariant surface condition \beq
\tfr32\sqrt{2}\,(u+b/3)u_x + u_t + \tfr32(u^3+bu^2+cu+d)=0.
\label{isc:512} \eeq Using this to eliminate $u_t$ in (\ref{pde:512})
and making the linearizing transformation $u=\sqrt{2}\,\Psi_x/\Psi$
yields  \beq 2\sqrt{2}\,\Psi_{xxx} + 2b\Psi_{xx} +\sqrt{2}\, c\Psi_x +
d \Psi=0. \label{eq:psi}\eeq The roots of \beq 2\sqrt{2}q^3 + 2 bq^2 +
\sqrt{2}\, cq + d =0\label{cubic:q} \eeq are $\tfr12\sqrt{2}\,m_1$,
$\tfr12\sqrt{2}\,m_2$ and $\tfr12\sqrt{2}\,m_3$,  where $m_1$, $m_2$
and $m_3$ are the roots of the cubic (\ref{cubic:u}).

In the following four subsections we consider the cases when $m_1$,
$m_2$ and $m_3$ are real and distinct, complex, two are equal and all
three are equal, respectively. These exponential-type solutions of
(\ref{pde:512}) are real for real $x$ and $t$ whereas those of
(\ref{pde:511}) are complex for real $x$ and $t$. Consequently we only
give details for (\ref{pde:512}) and leave it for the reader to derive
the analogous solutions for (\ref{pde:511}) using the transformation
$x\to\pm\i x$, $t\to -t$.

\subsubsection{Distinct Real Roots} Solving (\ref{eq:psi}) in the case
when $m_1$, $m_2$ and $m_3$ are distinct and real yields \beq \Psi(x,t)
= \mu_1(t)\exp\left(\tfr12\sqrt{2}\,m_1 x\right)+
\mu_2(t)\exp\left(\tfr12\sqrt{2}\,m_2 x\right)+
\mu_3(t)\exp\left(\tfr12\sqrt{2}\,m_3 x\right) \eeq where $\mu_1(t)$,
$\mu_2(t)$ and $\mu_3(t)$ are to be determined. By requiring that
$u=\sqrt{2}\,\Phi_x/\Phi$ also satisfies (\ref{isc:512}), or
equivalently  (\ref{pde:512}), it is easily shown that \beqn \mu_j(t) =
\cc{j} \exp\left\{ m_j\left(m_1+m_2+m_3-\tfr32m_j\right)t\right\},
\qquad j=1,2,3,\eeqn where $\cc1$, $\cc2$ and $\cc3$ are arbitrary
constants (note that $m_1+m_2+m_3=-b$).  Hence we have obtained the
following exact solution of (\ref{pde:512}) given by \beq u(x,t) =
{\cc1m_1\Psi_1(x,t)+\cc2m_2\Psi_2(x,t)+\cc3m_3\Psi_3(x,t)
\over\cc1\Psi_1(x,t)+\cc2\Psi_2(x,t)+\cc3\Psi_3(x,t)}
\label{sol:512}\eeq where \beqn\Psi_j(x,t) =
\exp\left\{\tfr12\sqrt{2}\,m_j x -
m_j\left(m_1+m_2+m_3-\tfr32m_j\right)t \right\}, \qquad j=1,2,3.  \eeqn
Setting $m_1=a$, $m_2=1$ and $m_3=0$ in (\ref{sol:512}), which we may
without loss of generality for the general cubic with real and distinct
roots, yields the following exact solution of the Fitzhugh-Nagumo
equation (\ref{fitznag}) for $a\not=0$ and $a\not=1$ \beq u(x,t)=
{a\cc1\exp\left\{\tfr12\left(\pm\sqrt2\,ax+ a^{2}t\right)\right\}
+\cc2\exp\left\{\tfr12\left(\pm\sqrt2\,x+ t\right)\right\} \over
\cc1\exp\left\{\tfr12\left(\pm\sqrt2\,ax+ a^{2}t\right)\right\}
+\cc2\exp\left\{\tfr12\left(\pm\sqrt2\,x+
t\right)\right\}+\cc3\exp\left(a t\right)}.\label{sol:fn}\eeq This
solution was obtained by Vorob'ev \cite{Vor} (who calls the associated
symmetry a ``partial symmetry of the first type''), Kawahara and Tanaka
\cite{KT} (using Hirota's bi-linear method \cite{Hirota}) and Hereman
\cite{Hereman} (using the truncated Painlev\'e expansion method
\cite{NTZ,Weiss,WTC}). Plots of (\ref{sol:fn}), with
$\cc1=\cc2=\cc3=1$, for (i), $a=-0.5$, (ii), $a=0.4$, (iii), $a=0.7$,
and (iv), $a=1.5$ are given in Figure 1; these plots were drawn using
{\sc maple}.

\subsubsection{Complex Roots} Suppose in this case that the roots of
(\ref{cubic:u}) are $m$ (real) and $\alpha\pm\i\beta$.  Then setting
$m_1=m$, $m_2=\alpha+\i\beta$ and $m_3=\alpha-\i\beta$ in
(\ref{sol:512}) we obtain the exact solution of \beq u_t = u_{xx} - u^3
+ (m+2\alpha)u^2-(\alpha^2+\beta^2+2\alpha m)u + (\alpha^2+\beta^2)m
\eeq given by \beq u(x,t) = {\cc1m\ex{\tfr12\sqrt2\,\mu x +
\left(\tfr12\mu^2+\tfr32\beta^2\right)t}
+\cc2\left(\alpha^2+\beta^2\right)^{1/2}\cos\left(\tfr12\sqrt2\,\beta x
- \mu\beta t+\theta_0+\delta_0\right)\over \cc1\ex{\tfr12\sqrt2\,\mu x
+ \left(\tfr12\mu^2+\tfr32\beta^2\right)t} +
\cc2\cos\left(\tfr12\sqrt2\,\beta x - \mu\beta t+\delta_0\right)}
\label{sol:512c}\eeq respectively, where $\cc1$, $\cc2$ and $\delta_0$
are arbitrary constants, $\mu=m-\alpha$ and
$\theta_0=\tan^{-1}(\beta/\alpha)$.

Setting $m=\alpha=0$ and $\beta=1$ in (\ref{sol:512c}) yields the exact
solution of \beq u_t = u_{xx} - u(u^2+1) \eeq given by \beq u(x,t) =
{\cc2\sin\left(\tfr12\sqrt2\, x\right)\over\cc1\exp\left(\tfr32t\right)
+ \cc2\cos\left(\tfr12\sqrt2\, x\right)}. \label{sol:512c1} \eeq
Setting $m=0$ and $\alpha=\beta=1$ in (\ref{sol:512c}) yields the exact
solution of \beq u_t = u_{xx} - u(u^2-2u+2) \eeq given by \beq u(x,t) =
{\cc2\left[\cos\left(\tfr12\sqrt2\, x - t\right) +
\sin\left(\tfr12\sqrt2\, x -  t\right)\right]
\over\cc1\exp\left(\tfr12\sqrt2\, x+2t\right) +
\cc2\cos\left(\tfr12\sqrt2\, x - t\right)}.\label{sol:512c2}\eeq Plots
of (\ref{sol:512c1}), with $\cc1=2$ and $\cc2=1$, and
(\ref{sol:512c2}), with $\cc1=1$ and $\cc2=1.1$, are given in Figure 2;
these plots were drawn using {\sc maple}.

\subsubsection{Two Equal Roots} Suppose that the cubic (\ref{cubic:u})
has a single root $m_1$ and a double root $m_2$. Then the solution of
(\ref{eq:psi}) is given by \beq \Psi(x,t) =
\mu_1(t)\exp\left(\tfr12\sqrt{2}\,m_1 x\right)+
\left[\mu_2(t)+x\mu_3(t)\right]\exp\left(\tfr12\sqrt{2}\,m_2 x\right)
\eeq where $\mu_1(t)$, $\mu_2(t)$ and $\mu_3(t)$ are to be determined.
A similar procedure to that used in the previous subsection yields the
exact solution of \beq u_t = u_{xx} - (u-m_1)(u-m_2)^2 \eeq given by
\beq u(x,t) = -\,{m_1\cc1\ex{\tfr12\left(\sqrt{2}\,\beta x+\beta
^2t\right)} +\cc2 \left[m_2\left(x-\sqrt{2}\,\beta
t\right)+\sqrt{2}\right]
 \over\cc1\ex{\tfr12\left(\sqrt{2}\,\beta x+\beta ^2t\right)} +
 \cc2\left(x-\sqrt{2}\,\beta t\right)} \eeq where $\beta=m_2-m_1$ and
$\cc1$, $\cc2$ and $\cc3$ are arbitrary constants.

In particular, an exact solution of \beq u_t = u_{xx} - u^2(u+b) \eeq
is \beq u(x,t) = -\,{b\cc1\ex{\tfr12\left(\sqrt{2}\,bx+b^2t\right)}
+\sqrt{2}\,\cc2 \over\cc1\ex{\tfr12\left(\sqrt{2}\,bx+b^2t\right)} +
\cc2\left(x-\sqrt{2}\,bt\right)}.  \label{plot3} \eeq Plots of this
solution for $b=\pm1$, $\cc1=1$ and $\cc2=\tfr12$ are given in Figure
3; these plots were drawn using {\sc maple}.

\subsubsection{Three Equal Roots} In the case when the cubic
(\ref{cubic:u}) has three equal roots, a similar procedure to that used
in the previous three subsections yields the exact solution of \beq u_t
= u_{xx} - u^3 \eeq given by \beq u(x,t) = {\sqrt{2}\,(2x+\cc1) \over
x^2 +\cc1 x+6t + \cc2}.  \label{plot4} \eeq A plot of this solution for
$\cc1=\cc2=0$ is given in Figure 4; this plot was drawn using {\sc
maple}.

In Table 3, we list the nonclassical exponential-type solutions of
equation (1.1) derived in this section for various canonical choices of
$f(u)$.

 \begin{table} \caption{Exponential-type solutions of equation (1.1)}
\vspace{3pt} \begin{tabular}{|@{\ }c@{\ }|@{\ }c@{\ }|} \hline $f(u)$ &
$u(x,t)$ \\*[2mm] \hline $u(1-u)(u-a)$ &
$\displaystyle{a\cc1\exp\left\{\tfr12\left(\pm\sqrt2\,ax+
a^{2}t\right)\right\} +\cc2\exp\left\{\tfr12\left(\pm\sqrt2\,x+
t\right)\right\} \over \cc1\exp\left\{\tfr12\left(\pm\sqrt2\,ax+
a^{2}t\right)\right\} +\cc2\exp\left\{\tfr12\left(\pm\sqrt2\,x+
t\right)\right\}+\cc3\exp\left(a t\right)}$ \\*[2mm] $- u(u^2+1)$ &
$\displaystyle{\cc2\sin\left(\tfr12\sqrt2\,
x\right)\over\cc1\exp\left(\tfr32t\right) +
\cc2\cos\left(\tfr12\sqrt2\, x\right)}$ \\*[2mm] $- u(u^2-2u+2)$ &
$\displaystyle{\cc2\left[\cos\left(\tfr12\sqrt2\, x - t\right) +
\sin\left(\tfr12\sqrt2\, x -  t\right)\right]
\over\cc1\exp\left(\tfr12\sqrt2\, x+2t\right) +
\cc2\cos\left(\tfr12\sqrt2\, x - t\right)}$ \\*[2mm] $-u^2(u+b)$ &
$\displaystyle-\,{b\cc1\ex{\tfr12\left(\sqrt{2}\,bx+b^2t\right)}
+\sqrt{2}\,\cc2 \over\cc1\ex{\tfr12\left(\sqrt{2}\,bx+b^2t\right)} +
\cc2\left(x-\sqrt{2}\,bt\right)}$\\*[2mm] $-u^3$ &
$\displaystyle{\sqrt{2}\,(2x+\cc1) \over x^2 +\cc1 x+6t + \cc2}$
\\*[2mm] \hline \end{tabular}

\vspace{12pt} \end{table} \subsection{Elliptic Function Solutions}
\def\sn{\mathop{\mbox{\rm sn}}\nolimits} \def\cn{\mathop{\mbox{\rm
cn}}\nolimits} \def\dn{\mathop{\mbox{\rm dn}}\nolimits}
\def\ds{\mathop{\mbox{\rm ds}}\nolimits} \def\sd{\mathop{\mbox{\rm
sd}}\nolimits}

In Case 4.1.1iii above we derived the nonclassical symmetry reduction
of \beq u_t = u_{xx} + u^3 \label{eq:52i}\eeq given by \beq u(x,t) =
(x+\cc1)w(z),\qquad z=\tfr12 x^2 + \cc1 x + 3t \label{sr:52i}\eeq where
$w(z)$ satisfies \beq w'' + w^3=0. \label{eq:52is}\eeq The solution of
this equation is \beq w(z) =
\tfr12\sqrt{2}\,\sd\left(z;\tfr12\sqrt{2}\right) \eeq where $\sd(z;k)$
is the Jacobi elliptic function satisfying \beq \left(\d\eta\over\d
z\right)^2= 1+(2k^2-1)\eta^2 + k^2(k^2-1)\eta^4.\label{jef:sd}\eeq
Hence we obtain the exact solution of (\ref{eq:52i}) given by \beq
u(x,t) = \tfr12\sqrt{2}\,(x+\cc1)\sd\left(\tfr12 x^2 + \cc1 x +
3t;\tfr12\sqrt{2}\right).  \label{plot5}\eeq A plot of this solution
for $\cc1=0$ is given in Figure 5; this plot was drawn using {\sc
mathematica}.

The analogous nonclassical symmetry reduction of \beq u_t = u_{xx} -
u^3 \label{eq:52ii}\eeq is given by \beq u(x,t) = (x+\cc1)w(z),\qquad
z=\tfr12 x^2 + \cc1 x + 3t \label{sr:52ii}\eeq where $w(z)$ satisfies
\beq w'' - w^3=0. \label{eq:52iis}\eeq The solution of this equation is
\beq w(z) = \sqrt{2}\,\ds\left(z;\tfr12\sqrt{2}\right) \eeq where
$\ds(z;k)$ is the Jacobi elliptic function satisfying \beq
\left(\d\eta\over\d z\right)^2= k^2(k^2-1)+(2k^2-1)\eta^2 +
\eta^4.\label{jef:ds}\eeq Hence we obtain the exact solution of
(\ref{eq:52ii}) given by \beq u(x,t) =
\sqrt{2}\,(x+\cc1)\ds\left(\tfr12 x^2 + \cc1 x +
3t;\tfr12\sqrt{2}\right).\eeq

In Case 4.1.1iv above we derived the nonclassical symmetry reduction of
\beq u_t = u_{xx} + u^3 +bu^2 + cu + {b(9c-2b^2)\over27}
\label{eq:52iii}\eeq given by \beq u(x,t) = \cc1\mu \sin(\mu x+\cc2)
\exp\left(-3\mu^2t\right)w(z) - \tfr13b, \quad z=\cc1\cos(\mu
x+\cc2)\exp\left(-3\mu^2t\right) \label{sr:52iii}\eeq where
$\mu^2=\tfr16(b^2-3c)$, $\cc1$ and $\cc2$ are arbitrary constants and
$w(z)$ satisfies (\ref{eq:52is}). Hence for $b^2>3c$ we obtain the
exact solution of (\ref{eq:52iii}) given by \beq u(x,t)
=\tfr12\sqrt{2}\,\cc1\mu \sin(\mu x+\cc2) \exp\left(-3\mu^2t\right)
\sd\left[\cc1\cos(\mu
x+\cc2)\exp\left(-3\mu^2t\right);\tfr12\sqrt{2}\right] - \tfr13b
\label{plot6}\eeq where $\mu^2=\tfr16(b^2-3c)$.  Two plots of this
solution for $\cc1=3$, $\cc2=0$ and $\mu=\tfr12$ are given in Figure 6;
these plots were drawn using {\sc mathematica}.  For $b^2<3c$ we obtain
the exact solution of (\ref{eq:52iii}) given by \beq u(x,t)
=\tfr12\sqrt{2}\,\cc1\lambda \sinh(\lambda x+\cc2)
\exp\left(3\lambda^2t\right)\sd\left[\cc1\cosh(\lambda
x+\cc2)\exp\left(3\lambda^2t\right);\tfr12\sqrt{2}\right] - \tfr13b
\label{plot7}\eeq where $\lambda^2=\tfr16(3c-b^2)$.  A plot of this
solution for $\cc1=3$, $\cc2=0$ and $\lambda=\tfr12$ is given in Figure
7; this plot was drawn using {\sc mathematica}.

The analogous nonclassical symmetry reduction of \beq u_t = u_{xx} -
u^3 -bu^2 - cu - {b(9c-2b^2)\over27} \label{eq:52iv}\eeq is given by
\beq u(x,t) = \cc1\mu \sinh(\mu x+\cc2) \exp\left(3\mu^2t\right)w(z)-
\tfr13b, \qquad z=\cosh(\mu x+\cc2)\exp\left(3\mu^2t\right) \eeq where
$\mu^2=\tfr16(b^2-3c)$, $\cc1$ and $\cc2$ are arbitrary constants and
$w(z)$ satisfies (\ref{eq:52iis}). Hence for $b^2>3c$ we obtain the
exact solution of (\ref{eq:52iv}) given by \beq    u(x,t) =
\sqrt{2}\,\cc1\mu \sinh(\mu x+\cc2) \exp\left(3\mu^2t\right)
\ds\left[\cc1\cosh(\mu
x+\cc2)\exp\left(3\mu^2t\right);\tfr12\sqrt{2}\right] - \tfr13b \eeq
and for $b^2<3c$, obtain the exact solution \beq u(x,t)
=\sqrt{2}\,\cc1\lambda \sin(\lambda x+\cc2)
\exp\left(-3\lambda^2t\right)\ds\left[\cc1\cos(\lambda
x+\cc2)\exp\left(-3\lambda^2t\right);\tfr12\sqrt{2}\right] - \tfr13b
\eeq where $\lambda^2=\tfr16(3c-b^2)$.

In Table 4, we list the nonclassical elliptic function solutions of
equation (1.1) derived in this section for various canonical choices of
$f(u)$.

 \begin{table} \caption{Elliptic Function solutions of equation (1.1)}
\vspace{3pt} \begin{tabular}{|@{\ }c@{\ }|@{\ }c@{\ }|} \hline $f(u)$ &
$u(x,t)$ \\*[2mm] \hline $u^3$ & $\displaystyle
\tfr12\sqrt{2}\,(x+\cc1)\sd\left(\tfr12 x^2 + \cc1 x +
3t;\tfr12\sqrt{2}\right)$ \\*[2mm] $-u^3$ & $\displaystyle
\sqrt{2}\,(x+\cc1)\ds\left(\tfr12 x^2 + \cc1 x +
3t;\tfr12\sqrt{2}\right)$ \\*[2mm] $u(u^2-2)$ & $\displaystyle
\tfr12\sqrt{2}\,\cc1\sin(x+\cc2) \exp\left(-3t\right)
\sd\left[\cc1\cos(x+\cc2)\exp\left(-3t\right);\tfr12\sqrt{2}\right]$
\\*[2mm] $u(u^2+2)$ & $\displaystyle \tfr12\sqrt{2}\,\cc1\sinh(x+\cc2)
\exp\left(3t\right)
\sd\left[\cc1\cosh(x+\cc2)\exp\left(3t\right);\tfr12\sqrt{2}\right]$
\\*[2mm] $-u(u^2-2)$ & $\displaystyle \sqrt{2}\,\cc1\sinh(x+\cc2)
\exp\left(3t\right)
\ds\left[\cc1\cosh(x+\cc2)\exp\left(3t\right);\tfr12\sqrt{2}\right]$
\\*[2mm] $-u(u^2+2)$ & $\displaystyle \sqrt{2}\,\cc1\sin(x+\cc2)
\exp\left(-3t\right)
\ds\left[\cc1\cos(x+\cc2)\exp\left(-3t\right);\tfr12\sqrt{2}\right]$
\\*[2mm] \hline \end{tabular}

\vspace{12pt} \end{table}

\newsection{Discussion} In this paper we have demonstrated that the
method of \DGBS\ has enabled all symmetries of the nonlinear heat
equation (1.1) to be found;  the list it provides of possible analytic
functions $f(u)$ for which symmetries may exist is definitive. The use
of \DGBS\ has made the analysis of overdetermined systems of
\PDES\ more tractable and whilst the diffgrob2 package needs to be used
interactively at present, nevertheless it is effective in solving the
overdetermined systems of determining equations for classical and
nonclassical symmetries of (1.1).

It is not clear how the direct method, developed by Clarkson and
Kruskal \cite{CK} for finding symmetry reductions of \PDES\ may be
applied to equations such as (1.1) which contain arbitrary functions.
Nucci and Clarkson \cite{NC} (see also \cite{CM}) demonstrated that the
symmetry reduction ({\ref{sol:fn}) of the Fitzhugh-Nagumo equation
(\ref{fitznag}), is obtainable using the nonclassical method due to
Bluman and Cole \cite{BCa}, though {\it not\/} using the direct method
due to Clarkson and Kruskal \cite{CK}. Hence the nonclassical method is
more general than the direct method, at least as it was originally
formulated. Olver \cite{Olverb} has recently shown that the direct
method is equivalent to the nonclassical method when the infinitesimals
$\xi$ and $\tau$ are independent of the dependent variable $u$ (see
also \cite{CH,Pucci,Zida,Zidb}). In this case the associated vector
field is of the form  $${\bf w} = \xi(x,t)\partial_x +
\tau(x,t)\partial_t + \phi(x,t,u)\partial_u,$$ and generates a group of
``{\it fibre-preserving transformations\/}'', meaning that the
transformations in $x$ and $t$ do not depend upon $u$. A recent
extension of the direct method by Estevez \cite{Este} does yield the
symmetry reduction solution ({\ref{sol:fn}) of the Fitzhugh-Nagumo
equation (\ref{fitznag}). These results pose the following important
open question:  ``for which \PDES\ does the nonclassical method yield
more symmetry reductions than the direct method?'' Furthermore, it
remains an open question to determine {\it a priori\/} which \PDES\
possess symmetry reductions that are not obtainable using the classical
Lie group approach.

Fushchich and Serov \cite{FushS} (see also \cite{Fush}) claimed that
\beq u_t + u_{xx} = f(u) \label{fush51} \eeq which is equivalent to
(1.1), has a nonclassical symmetry, which they refer to as a
conditional symmetry with respect to the vector field (1.3), if and
only if (1.1) is equivalent to the special form  \beq u_t + u_{xx} = a
u^3+b u+c \label{gcubic51} \eeq where $a$, $b$ and $c$ are arbitrary
constants, which is equivalent to (5.1) and (5.2) and that the
associated infinitesimals are \beq \xi = \tfr23\sqrt{2a}\,u,\qquad
\tau=1,\qquad \phi=\tfr32 (a u^3+b u+c). \label{fush:isc}\eeq
Subsequently they consider the four special cases of (\ref{gcubic51})
given by \bear u_t + u_{xx} &=& a u(u^2-1), \label{fuchi} \\ u_t +
u_{xx} &=& a u(u^2+1), \label{fuchii} \\ u_t + u_{xx} &=& a (u^3-3u+2),
\label{fuchiii} \\ u_t + u_{xx} &=& a u^3 \label{fuchiv} \eear and
whereas they integrate the invariant surface conditions, they do not
write down the associated exact solutions of (\ref{fush51}).

The results we have obtained in this paper show that there are more
nonclassical symmetry of (\ref{fush51}) other than those generated by
the infinitesimals (\ref{fush:isc}). In particular they do not obtain
the nonclassical symmetry reductions corresponding to those obtained in
Cases 4.1.1iii,iv and 4.2.3ii,iii above. Further even for the symmetry
reductions associated with (\ref{fush:isc}), the four  ``canonical''
equations (\ref{fuchi},\ref{fuchii},\ref{fuchiii},\ref{fuchiv}) do not
include the case when the roots of the cubic $a u^3+b u+c$ are real,
distinct and unequally spaced.

\section*{Acknowledgements} We thank Clara Nucci and Peter Olver for
helpful discussions and the referees for constructive comments. We also
thank the Program in Applied Mathematics, University of Colorado at
Boulder, for their hospitality, where this work was completed. The
support of SERC (grant GR/H39420) is gratefully acknowledged.  PAC is
also grateful for support through a Nuffield Foundation Science
Fellowship and NATO grant CRG 910729.

\section*{Figure Captions}

\smallskip\noindent{\bf Figure 1}:\quad  Exponential solution
(\ref{sol:fn}) of $u_t=u_{xx}+u(1-u)(u-a)$ where (i) $a=-0.5$, (ii)
$a=0.4$, (iii)  $a=0.7$, (iv)  $a=1.5$.

\smallskip\noindent{\bf Figure 2a}:\quad  Exponential solution
(\ref{sol:512c1}) of $u_t=u_{xx}-u(u^2+1)$ where $\cc1=\cc2=1$.

\smallskip\noindent{\bf Figure 2b}:\quad  Exponential solution
(\ref{sol:512c2}) of $u_t=u_{xx}-u(u^2-2u+2)$ where $\cc1=\cc2=1$.

\smallskip\noindent{\bf Figure 3}:\quad  Exponential solution
(\ref{plot3}) of $u_t=u_{xx}-u(u^2+b)$ where (i) $b=-1$ and (ii)
$b=1$.

\smallskip\noindent{\bf Figure 4}:\quad  Rational solution
(\ref{plot4}) of $u_t=u_{xx}+u^3$ where $\cc1=\cc2=0$.

\smallskip\noindent{\bf Figure 5}:\quad  Elliptic solution
(\ref{plot5}) of $u_t=u_{xx}+u^3$ where $\cc1=0$.

\smallskip\noindent{\bf Figure 6}:\quad  Elliptic solution
(\ref{plot6}) of $u_t=u_{xx}+u^3+cu+{b(9c-2b^2)/27}$ where
$\mu^2=\tfr16(b^2-3c)=\tfr12\sqrt{2}$, $\cc1=3$ and $\cc2=0$.

\smallskip\noindent{\bf Figure 7}:\quad  Elliptic solution
(\ref{plot7}) of $u_t=u_{xx}+u^3+cu+{b(9c-2b^2)/27}$ where
$\lambda^2=\tfr16(3c-b^2)=\tfr12\sqrt{2}$, $\cc1=3$ and $\cc2=0$.

\end{document}